\documentclass[10pt,twocolumn,letterpaper]{article}

\usepackage{iccv}
\usepackage{times}
\usepackage{epsfig}
\usepackage{graphicx}
\usepackage{amsmath}
\usepackage{amssymb}

\usepackage{enumitem}
\usepackage[ruled,vlined]{algorithm2e}
\usepackage{subfigure}
\usepackage[accsupp]{axessibility}  % Improves PDF readability for those with disabilities.

\usepackage[breaklinks=true,bookmarks=false]{hyperref}

\iccvfinalcopy % *** Uncomment this line for the final submission

\def\iccvPaperID{2201} % *** Enter the ICCV Paper ID here

\ificcvfinal\fi

\def\etal{\emph{et al. }}

\renewcommand\v[1]{\mathbf{#1}}

\newcommand\xt{\v{x}_t}
\newcommand\xhatt{\widehat{\v{x}}_{t}}
\newcommand\xtildet{\widetilde{\v{x}}_{t}}

\newcommand\mhat{\widehat{\v{m}}}

\newcommand\fhatt{\widehat{\v{f}}_{t}}

\newcommand\fzeroone{\v{f}_{0\rightarrow1}}
\newcommand\fhattzero{\widehat{\v{f}}_{t\rightarrow0}}
\newcommand\ftildetzero{\widetilde{\v{f}}_{t\rightarrow0}}
\newcommand\fonezero{\v{f}_{1\rightarrow0}}
\newcommand\fhattone{\widehat{\v{f}}_{t\rightarrow1}}
\newcommand\ftildetone{\widetilde{\v{f}}_{t\rightarrow1}}

\newcommand\rt{\v{r}_t}
\newcommand\rhatt{\widehat{\v{r}}_{t}}

\newcommand\xref{\widehat{\v{x}}_{\textup{ref}}}
\newcommand\xrefzero{\widehat{\v{x}}_{\textup{ref}_0}}
\newcommand\xrefone{\widehat{\v{x}}_{\textup{ref}_1}}

\newcommand{\ssf}{\operatorname{SSF}}
\newcommand{\iframeae}{\operatorname{Image-AE}}
\newcommand{\flowae}{\operatorname{Flow-AE}}
\newcommand{\residualae}{\operatorname{Residual-AE}}

\newcommand{\superslomo}{\operatorname{Super-SloMo}}
\newcommand{\flownet}{\operatorname{FlowNet}}
\newcommand{\refinenet}{\operatorname{RefineNet}}
\newcommand{\flowinterp}{\operatorname{Flow Interpolation}}
\newcommand{\bidirwarp}{\operatorname{BidirWarp}}

\newcommand{\warp}{\operatorname{Warp}}

\newcommand{\ibi}{\operatorname{IBI}}
\newcommand{\ibp}{\operatorname{IBP}}

\newcommand{\hier}{\operatorname{hierarchical}}
\newcommand{\seq}{\operatorname{sequential}}

\newcommand{\avc}{\operatorname{H.264}}
\newcommand{\hevc}{\operatorname{H.265}}

\newcommand{\rate}{\operatorname{Rate}}
\newcommand{\mse}{\operatorname{MSE}}
\newcommand{\psnr}{\operatorname{PSNR}}
\newcommand{\msssim}{\operatorname{MS-SSIM}}

\newcommand{\bepic}{\operatorname{B-EPIC}}
\newcommand{\bepicmse}{\operatorname{B-EPIC(MSE)}}
\newcommand{\bepicmsssim}{\operatorname{B-EPIC(MS-SSIM)}}

\newcommand{\ffmpeg}{\texttt{FFMPEG}}

\begin{document}

%%%%%%%%% TITLE
\title{Extending Neural P-frame Codecs for B-frame Coding}

% \author{Reza Pourreza\\
% Qualcomm AI Research\\
% San Diego CA\\
% {\tt\small pourreza@qti.qualcomm.com}
% \and
% Taco Cohen\\
% Qualcomm AI Research\\
% Amsterdam, Netherlands\\
% {\tt\small tacos@qti.qualcomm.com}
% }

\author{Reza Pourreza and Taco Cohen\\
Qualcomm AI Research\thanks{Qualcomm AI Research is an initiative of Qualcomm Technologies, Inc.} \\
{\tt\small \{pourreza,tacos\}@qti.qualcomm.com}
}

\maketitle
\ificcvfinal\thispagestyle{empty}\fi

%%%%%%%%% ABSTRACT
\begin{abstract}
While most neural video codecs address P-frame coding (predicting each frame from past ones),
in this paper we address B-frame compression (predicting frames using both past and future reference frames).
Our B-frame solution is based on the existing P-frame methods.
As a result, B-frame coding capability can easily be added to an existing neural codec.
%While P-frame codecs often use a single reference frame, usually the previous frame, to encode an input frame, B-frame codecs use multiple reference frames, usually a past and future one, to do so.
The basic idea of our B-frame coding method is to interpolate the two reference frames to generate a single reference frame and then use it together with an existing P-frame codec to encode the input B-frame.
Our studies show that the interpolated frame is a much better reference for the P-frame codec compared to using the previous frame as is usually done.
Our results show that using the proposed method with an existing P-frame codec can lead to 28.5\% saving in bit-rate on the UVG dataset compared to the P-frame codec while generating the same video quality.
\end{abstract}

%%%%%%%%% BODY TEXT
\section{Introduction}

There are two types of frames in the video coding domain, Intra-frames and Inter-frames. 
Intra-frames (I-frames) are encoded/decoded independently of other frames. I-frame coding is equivalent of image compression. Inter-frames are encoded using motion compensation followed by residuals \ie a prediction of an input frame is initially devised by moving pixels or blocks of one or multiple reference frames and then the prediction is corrected using residuals. Prediction is an essential task in inter-coding, for it is the primary way in which temporal redundancy is exploited.
%as that is the main component of codecs that models the temporal aspect of videos which is the main source of redundancy.
In the traditional paradigm of video coding~\cite{sullivanOverviewHighEfficiency2012,avc}, motion vectors are used to model the motion of blocks of pixels between a reference and an input image~\cite{sullivanOverviewHighEfficiency2012}. In the neural video coding domain, dense optical flow is usually used to model individual pixels movements. In both cases, a warping is performed on references using motion vectors or optical flow to generate the prediction.

\begin{figure}[tb]
  \begin{center}
    \includegraphics[width=1.\linewidth]{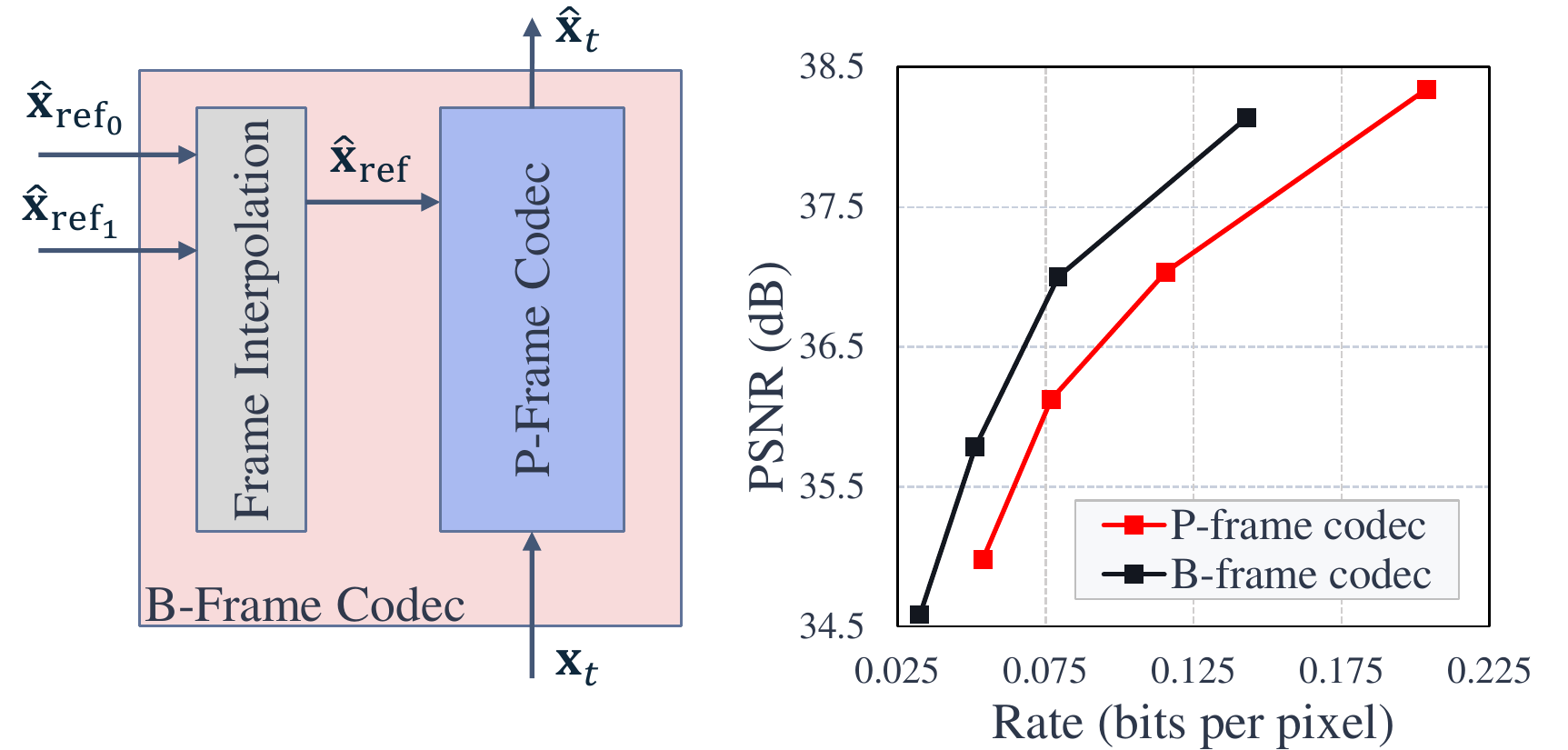}
  \end{center}
  \vspace{-1em}
  \caption{(a) the general idea of this work \ie extending an existing P-frame codec to a B-frame codec by adding an interpolation block, (b) the rate-distortion improvements on the UVG dataset~\cite{UVG} where the P-frame~\cite{Agustsson_2020_CVPR} and the B-frame codecs are trained on the Vimeo-90k dataset~\cite{Xue_2019} for the same number of iterations. The improvement is equivalent of 28.5\% saving in bit-rate measured by BD-rate gain~\cite{bdrate}.}
  \label{fig:intro}
  \vspace{-1.5em}
\end{figure}

\begin{figure*}[t]
  \begin{center}
   \includegraphics[width=1.\textwidth]{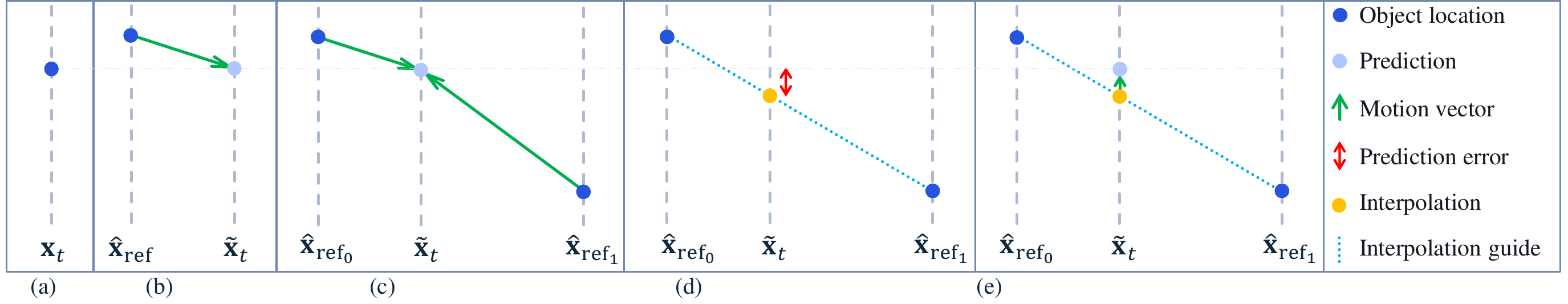}
  \end{center}
  \vspace{-1em}
   \caption{Prediction in inter-frame coding. $\xt$, $\xtildet$, and $\xref$ denote an input frame, the corresponding prediction and the reference, respectively.
   (a) Actual object location.
   (b) P-Frame prediction, a motion vector with respect to a single reference is transmitted. 
   (c) B-frame prediction based on bidirectional flow/warp, two motion vectors with respect to two references are transmitted. 
   (d) B-frame prediction based on frame interpolation, the interpolation result is treated as the prediction. No motion information is transmitted. 
   (e) Our B-frame prediction approach, the interpolation result is corrected using a unidirectional motion vector similar to P-frame.}
  \label{fig:idea}
  \vspace{-1em}
\end{figure*}

Inter-frames are further divided into Predicted (P) frames and Bi-directional predicted (B) frames. P-frame coding, which is suitable for low-latency applications such as video conferencing, uses only past decoded frames as references to generate a prediction. Most of the available literature on neural inter coding falls under this category and the methods often use a single past decoded frame as reference~\cite{luDVCEndtoendDeep2019} (see Fig.~\ref{fig:idea}.b). On the other hand, B-frame coding, which is suitable for applications such as on-demand video streaming, uses both past and future decoded frames as references. Future references provide rich motion information that facilitate frame prediction and eventually lead to better coding efficiency. The number of neural video codecs that address B-frame coding is limited~\cite{Cheng_2019_CVPR,Djelouah_2019_ICCV,Habibian_2019_ICCV,wuVideoCompressionImage2018}. They use two references and generate a prediction either by bidirectional optical flow estimation and warping or by performing frame interpolation. The reported results show that these approaches, despite relative success in video coding, do not fully exploit the motion information provided by two references as the results are not competitive with state-of-the-art P-frame codecs~\cite{Agustsson_2020_CVPR}.

For a given input frame, when references from both past and future are available, under a linear motion assumption, one can come up with a rough prediction of the input frame by linearly interpolating the two references. 
%This prediction consumes zero bits since the two references are already available and no motion field or other information is required to interpolate.
This prediction does not need to be coded since the two references are already available to the receiver.
%{\color{red} As a result, no motion field or other information needs to be transmitted as part of the bit-stream to interpolate.}
%and comes for free in terms of bandwidth consumption.
The neural B-frame coding methods that work based on bidirectional flow/warping~\cite{Djelouah_2019_ICCV}, do not use this useful information and send the optical flows with respect to both references (see Fig.~\ref{fig:idea}.c). On the other hand, the interpolation outcome is only accurate under linear motion assumption. So in the neural B-frame models that rely on frame interpolation~\cite{Cheng_2019_CVPR,wuVideoCompressionImage2018}, the prediction is likely to not exactly be aligned with the input frame (see Fig.~\ref{fig:idea}.d). Even when a  non-linear frame interpolator is employed~\cite{quadratic_NEURIPS2019_d045c59a}, misalignment could still occur. In these situations, the codec solely relies on residuals to compensate for the misalignment. As a result, coding efficiency could be significantly lower compared to a scenario where the misalignment is mitigated via some inexpensive side-information first before applying residual coding.

In this work, we address this issue by introducing a new approach for neural B-frame coding, which despite its simplicity, is proven very effective.
The method involves interpolating two reference frames to obtain a single reference frame, which is then used by a P-frame model to predict the current frame (see Fig.~\ref{fig:intro} and Fig.~\ref{fig:idea}.e).
A residual is applied to this prediction.
%We employ frame interpolation and a P-frame codec to construct our B-frame codec where frame interpolation is used to interpolate the available references to generate a single reference frame and the P-frame codec initially corrects the misalignments between the interpolation outcome and the input frame by motion compensation and then corrects the prediction using residuals.

Our method takes advantage of the rich motion information available to the receiver by performing frame interpolation and does not suffer from the residual penalty due to misalignment. Since our B-frame coding solution operates based on a P-frame codec, an existing P-frame codec can be used to code B-frames.
In fact, the same network can learn to do both P-frame coding as well as contributing to B-frame compression. In other words, by adding a frame interpolator to a P-frame codec, the codec is able to code both P-frames and B-frames. One can freely choose an existing interpolation and P-frame method when implementing our technique. %our solution works with most of the existing frame interpolation and P-frame coding methods.

In video coding, videos are split into groups of pictures (GoP) for coding. The neural video codec that we develop in this work $\bepic$ (B-Frame compression through Extended P-frame \& Interpolation Codec) supports all frame types. Given that different frame types yield different coding efficiencies, it is crucial to choose the right frame type for the individual frames in a GoP. In this work, we look closely into GoP structure.

Our main contributions and findings are as follows:
\begin{itemize}[leftmargin=*]
    \vspace{-.8em}
    \setlength\itemsep{-.4em}
    \item We introduce a novel B-frame coding approach based on existing P-frame codecs and frame interpolation,
    \item A single P-frame network is used for both P-frame and B-frame coding through weight-sharing,
    \item A thorough analysis of the effect of GoP structure on performance is provided,
    \item The proposed solution outperforms existing neural video codecs by a significant margin and achieves new state-of-the-art results.
\end{itemize}

\begin{figure*}[tb]
  \begin{center}
    \includegraphics[width=1.\textwidth]{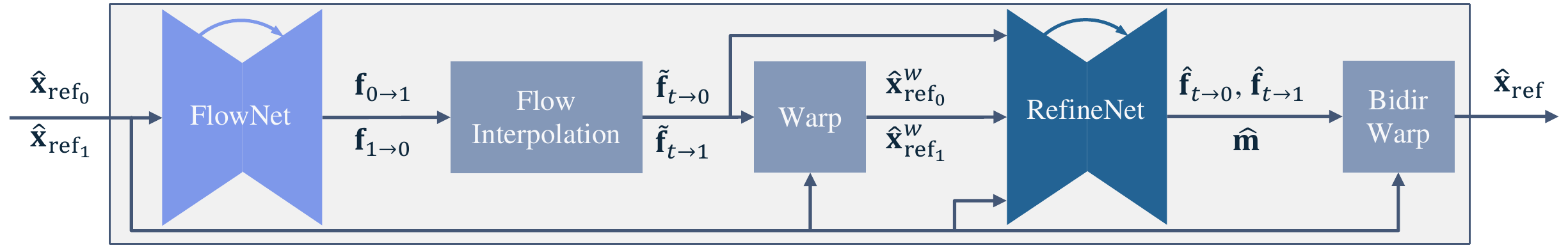}
  \end{center}
  \vspace{-1em}
  \caption{Frame interpolation method based on $\superslomo$~\cite{superslomo}. $\flownet$ and $\refinenet$ are trainable while $\flowinterp$, $\warp$, and $\bidirwarp$ are non-trainable operators. See section \ref{sec:interpolation} for details.}
  \label{fig:superslomo}
  \vspace{-1em}
\end{figure*}

\section{Related work}

\textbf{I-frame/Image coding:}
Great progress has been made in the development of neural image codecs. 
%There has been great progress in the development of neural image codecs.
Research has focused on various aspects of neural coding, such as 
%Existing methods span a wide range of techniques that address different aspects of neural coding such as
architecture~\cite{balleDensity2015,mentzerConditionalProbabilityModels2018,rippel2017real,todericiFullResolutionImage2017}, quantization~\cite{agustssonSofttoHardVectorQuantization2017}, priors~\cite{balleVARIATIONALIMAGECOMPRESSION2018,mentzerConditionalProbabilityModels2018}, and multi-rate coding~\cite{Choi_2019_ICCV,lu2021progressive,todericiVariableRateImage2015}. Recently, a hierarchical hyperprior model~\cite{balleEndtoendOptimizationNonlinear2016, balleVARIATIONALIMAGECOMPRESSION2018} has been widely adopted in the neural coding field and there are multiple variants including some equipped with autoregressive models~\cite{minnenJointAutoregressiveAndHierarchicalPriors2018,minnen2020channelwise} and attention mechanisms~\cite{liuNLAICImageCompression2019}.

\textbf{P-frame coding:} 
Most of the existing neural video codecs fall under this category where unidirectional motion estimation/compensation is followed by residual correction~\cite{liu2020learned,stephan_2019_neurips,rippelLearnedVideoCompression2018}. Lu \etal introduced DVC~\cite{luDVCEndtoendDeep2019}, a basic P-frame codec which is later upgraded in~\cite{DVC_TPAMI}. While motion is often modelled using spatial optical flow, Agustsson \etal introduced scale-space flow~\cite{Agustsson_2020_CVPR} to address uncertainties in motion estimation via a blur field which is further enhanced in~\cite{yang2021hierarchical}. Recent works have introduced more sophisticated components, \eg Golinski \etal~\cite{Golinski_2020_ACCV} added re currency to capture longer frame dependencies, Lin \etal look at multiple previous frames to generate a prediction in M-LVC~\cite{mlvc}, Liu \etal perform multi-scale warping in feature space in NVC~\cite{9247134}, and Chen \etal~\cite{chen2020learning} replaced optical flow and warping by displaced frame differences.

\textbf{B-frame coding:} 
Wu \etal~\cite{wuVideoCompressionImage2018} introduced one of the pioneering neural video codecs via frame interpolation that was facilitated by context information. Chang \etal~\cite{Cheng_2019_CVPR} improved the idea through adding a residual correction. Habibian \etal~\cite{Habibian_2019_ICCV} provided an implicit multi-frame coding solution based on 3D convolutions. Djelouah \etal~\cite{Djelouah_2019_ICCV} employed bidirectional optical flow and warping feature domain residuals for B-frame coding. A recent work~\cite{9300040} provides a multi-reference video codec that could be applied to both P-frame and B-frame coding.

% It is worth to mention that traditional videos codecs such as $\avc$~\cite{avc} and $\hevc$~\cite{sullivanOverviewHighEfficiency2012}, support all these frame-types and can be configured to code videos using combinations of them.

\section{Method}
% In this section we provide more details about the developed method.

\begin{figure}[t!]
  \begin{center}
    \includegraphics[width=1.0\linewidth]{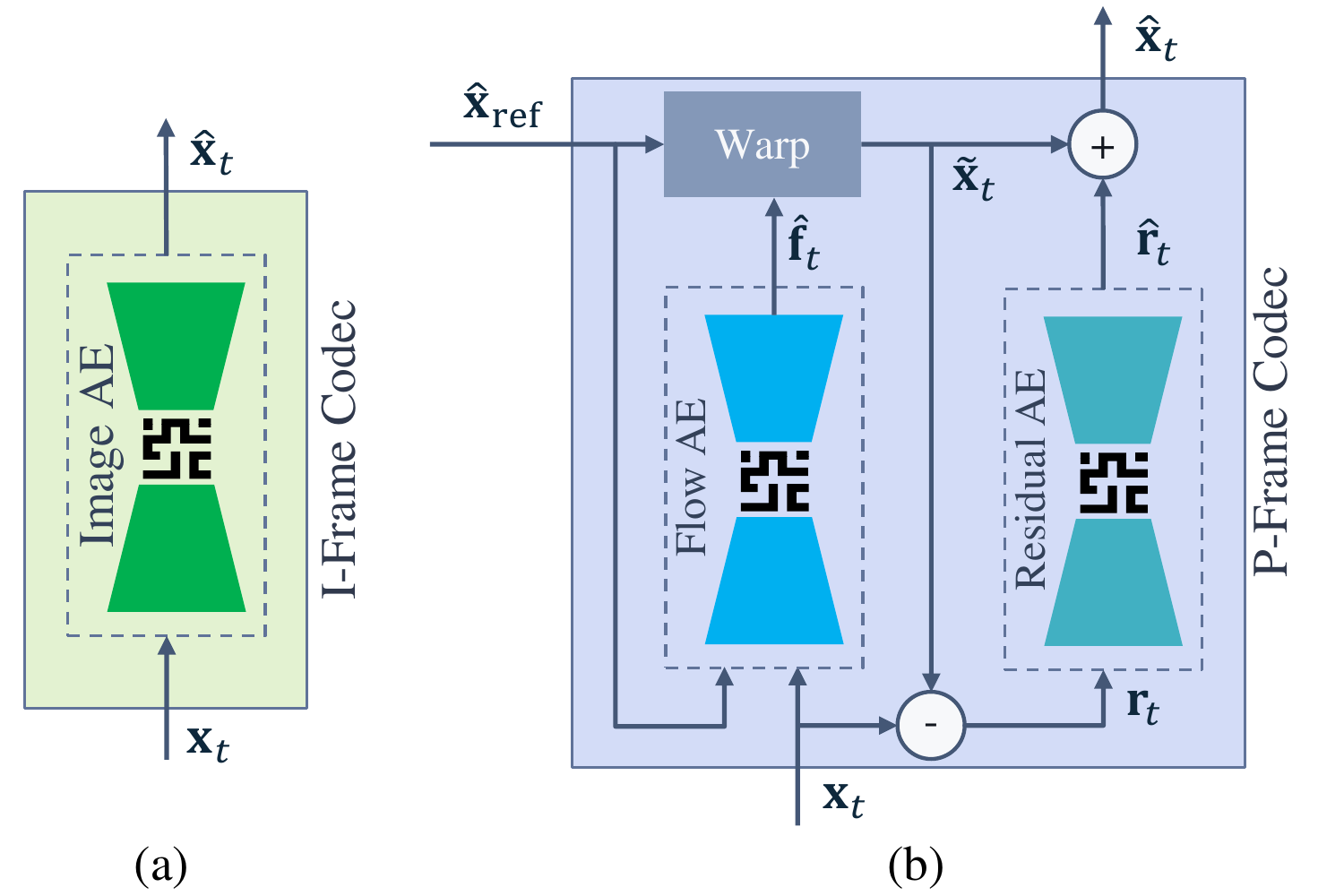}
  \end{center}
  \vspace{-1em}
  \caption{Block diagrams of (a) I-frame and (b) P-frame codecs, both based on $\ssf$~\cite{Agustsson_2020_CVPR}. See section~\ref{sec:i_p_codec} for more details.}
  \label{fig:ip_codec}
  \vspace{-3em}
\end{figure}

%Our system 
We develop a neural video codec $\bepic$ that consists of an I-frame codec, a P-frame codec, and a frame interpolator.
%(see Fig. \ref{fig:intro} \& \ref{fig:ip_codec}).
The I-frame codec (Fig. \ref{fig:ip_codec}.a) encodes individual frames $\xt$ independently to produce a reconstruction $\xhatt$.
The P-frame codec applies a warp to a reference frame $\xref$
%(typically $\widehat{\mathbf{x}}_{t-1}$) 
to produce a prediction $\xtildet$ of $\xt$, which is then corrected by a residual to obtain the reconstruction $\xhatt$ (see Fig. \ref{fig:ip_codec}.b).
The frame interpolator takes two reference frames $\xrefzero$ and $\xrefone$ 
%(with $\textup{ref}_0 < t < \textup{ref}_1$) 
and produces an interpolation $\xref$ (see Fig.~\ref{fig:superslomo}).
%a prediction $\xtildet$ of $\xt$.

Our novel B-frame codec (Fig. \ref{fig:intro}) works by using the frame interpolator on references $\xrefzero$ and $\xrefone$ to produce a single reference $\xref$, which is then used by the P-frame codec to encode $\xt$.
The resulting system thus supports I-frames, P-frames and B-frames in a flexible manner.

Although our general method can be implemented using any frame interpolator and P-frame codec, in this work we develop a specific codec that uses the $\superslomo$~\cite{superslomo} frame interpolator and the Scale-Space Flow ($\ssf$) codec~\cite{Agustsson_2020_CVPR}.
%$\ssf$ is equipped with I- and P-frame codecs.
%We pair the $\ssf$ P-frame codec with $\superslomo$ to build a B-frame codec as well.
%It is worth to note that the 
The $\ssf$ P-codec is used within our video codec in a stand-alone fashion as well as in a B-frame codec when bundled with $\superslomo$, while the two instances share weights. 
%Overall, our video codec consists of three components, an I-frame codec, a P-frame codec, and a frame interpolator, explained next.
In the following subsections we discuss $\superslomo$ and $\ssf$, as well as the GoP structure and loss function in more detail.

\subsection{Frame interpolation}
\label{sec:interpolation}

In the frame interpolation block, the goal is to interpolate two references whose time indices are normalized to $0$ and $1$, \ie $\xrefzero$ and $\xrefone$, to time $t$ where $0<t<1$. Since in B-frame coding, $\xt$ could be anywhere between $\xrefzero$ and $\xrefone$, an important factor in choosing $\superslomo$ over other competitors is that $\superslomo$ supports an arbitrary $t: \:0<t<1$ while many other methods assume $t=0.5$. The latter can be still used within our model, though they will impose restrictions on the GoP size. See section~\ref{sec:gop_structure} for more details.
%{\color{red} Does this contradict our frame that we can use most/any interpolation method with our method?}

The block diagram of $\superslomo$ is depicted in Fig.~\ref{fig:superslomo}. $\superslomo$ consists of two trainable components \ie $\flownet$ and $\refinenet$ as well as non-trainable $\flowinterp$, $\warp$, and $\bidirwarp$ blocks. Optical flow between $\xrefzero$ and $\xrefone$ in the forward and backward directions, \ie $\fzeroone$ and $\fonezero$ where $\fzeroone$ denotes the optical flow from $\xrefzero$ to $\xrefone$, are initially calculated in $\flownet$ and then interpolated at time $t$ in $\flowinterp$ using linear interpolation:
\begin{align}
  \label{eq:interp}
  \begin{split}
    & \ftildetzero = -\big( 1-t \big) \: t \: \fzeroone + t^2 \: \fonezero
    \\
    & \ftildetone = \big( 1-t \big)^2 \: \fzeroone - t \: \big( 1-t \big) \: \fonezero
  \end{split}
\end{align}

$\xrefzero$ and $\xrefone$ are then warped using the interpolated optical-flows $\ftildetzero$ and $\ftildetone$. The two warped references together with the original references and the interpolated optical flows are given to $\refinenet$ for further adjustment of the bidirectional optical flows \ie $\fhattzero$, $\fhattone$, and generating a mask $\mhat$. The interpolation result is finally generated using bidirectional warping:
\begin{align}
  \label{eq:bidirwarp}
  \begin{split}
    \xref =& \warp \big( \xrefzero, \: \fhattzero \big ) \odot \mhat \:  +\\
    & \warp \big( \xrefone, \: \fhattone \big ) \odot \big( 1 - \mhat \big)
  \end{split}
\end{align}

\noindent where $\odot$ denotes element-wise multiplication.

In this work, $\flownet$ and $\refinenet$ are implemented using a $\operatorname{PWC-Net}$~\cite{Sun2018PWC-Net} and a $\operatorname{U-Net}$~\cite{UNet}, respectively. See Appendix~\ref{appendix:arch} for a more detailed illustration of the $\refinenet$ architecture.

\subsection{I-/P-frame codecs}
\label{sec:i_p_codec}

Our I-frame and P-frame codecs are depicted in Fig.~\ref{fig:ip_codec}. While the I-frame codec consists of a single autoencoder $\iframeae$ that compresses $\xt$ to a reconstruction $\xhatt$, the P-frame codec first generates a prediction $\xtildet$ of $\xt$ through motion estimation via $\flowae$ and motion compensation via $\warp$ and then corrects $\xtildet$ using residuals via $\residualae$ to reconstruct $\xhatt$

%The $\ssf$ P-frame codec consists of two components, \emph{i)} motion estimation/compensation and \emph{ii)} residual, where each component has a dedicated autoencoder \ie $\flowae$ and $\residualae$. Motion estimation is done via $\flowae$, motion compensation is done via $\warp$, and residual correction is done via $\residualae$ 
\begin{alignat}{2}
%\begin{align}
  \label{eq:p_codec}
%   \begin{split}
    \fhatt &= \flowae \big( \xt, \: \xref \big), \:\:\:\: &
    \xtildet &= \warp \big( \xref, \: \fhatt \big) \notag \\
    \rt &= \xt - \xtildet, &
    \rhatt &= \residualae \big( \rt \big) \\
    \xhatt &= \xtildet + \rhatt & \notag
%   \end{split}
%\end{align}
\end{alignat}

\noindent where $\fhatt$, $\rt$, and $\rhatt$ denote optical flow, encoder residual, and decoder residual, respectively.

\begin{figure}[!b]
  \vspace{-.7em}
  \begin{center}
    \includegraphics[width=.8\linewidth]{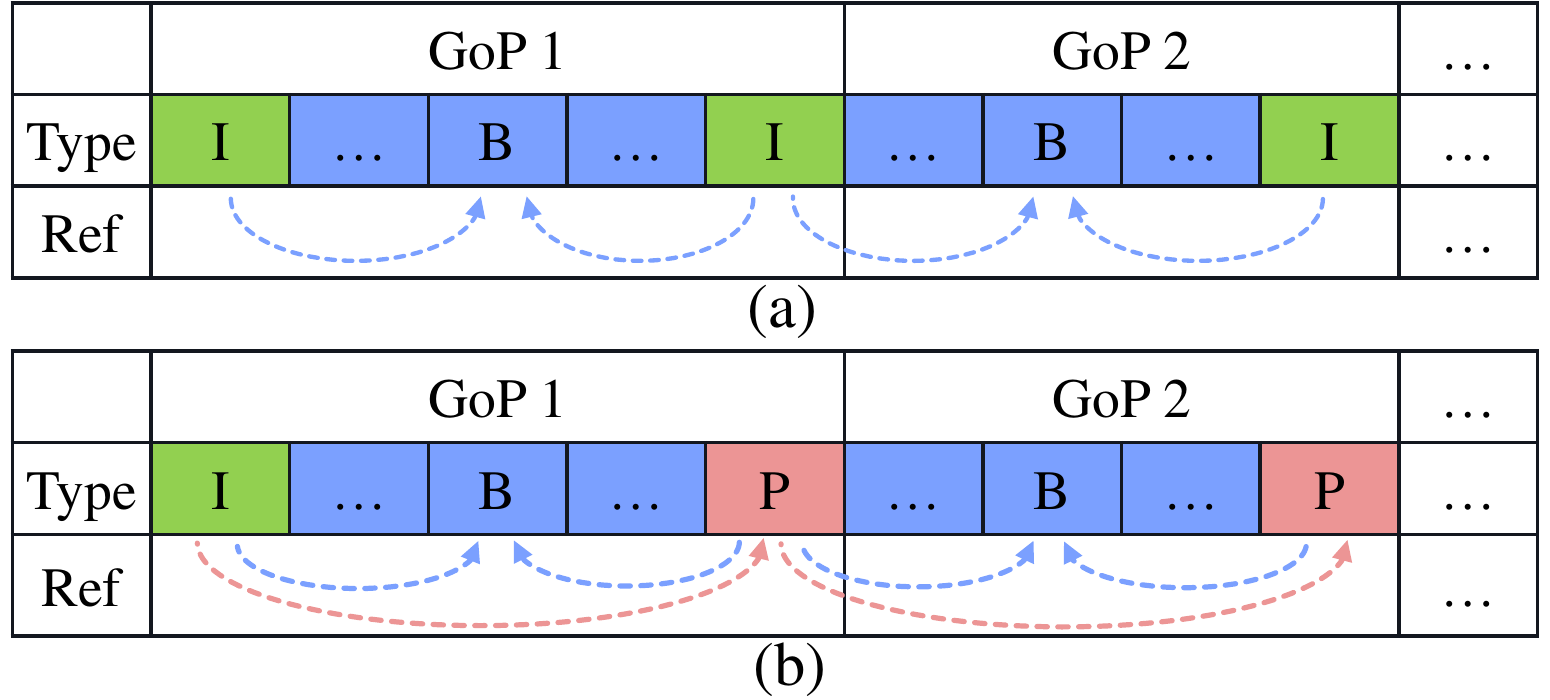}
  \end{center}
  \vspace{-1em}
  \caption[Caption for LOF]{Frame type selection\footnotemark, the first reference of a video is always coded as I-frame. (a) In the $\ibi$ configuration, the subsequent references are coded as I-frame as well while (b) in the $\ibp$ configuration, they are coded as P-frame. The arrows show the references used in inter-coding.}
  \label{fig:gop_struct}
%   \vspace{-2em}
\end{figure}

In $\ssf$, $\fhatt$ consists of spatial and scale displacement maps and $\warp$ is a trilinear interpolation operator on a blur stack of $\xref$. While $\ssf$ uses Gaussian filters to generate a blur stack and uses scale to \textit{non-linearly} point to the blur stack, we generate the blur stack using a Gaussian pyramid followed by bilinearly upsampling all pyramid scales to the original resolution and use scale to \textit{linearly} point to the blur stack.

% The $\ssf$ I-frame codec consists of a single autoencoder $\iframeae$. 
All the above autoencoders \ie $\iframeae$, $\flowae$, and $\residualae$, have the same architecture (without weight sharing) based on the mean-scale hyperprior model~\cite{balleVARIATIONALIMAGECOMPRESSION2018} that consists of a main autoencoder (an encoder and a decoder) and a hyperprior (a hyper-encoder, a mean hyper-decoder and a scale hyper-decoder) where all the components are parameterized via convolutional neural networks. The quantized latent variables $\mathbf{z}$ are broken down into a latent and a hyper-latent where the latent has a Gaussian prior whose probabilities are conditioned on the hyper-latent and the hyper-latent has a data-independent factorized prior.
%The mean-scale hyperprior model consists of an autoencoder where the encoder (analysis transform) and the decoder (synthesis transform) are parameterized via convolutional neural networks, and its latent variable $\mathbf{z}$ has a Gaussian prior whose mean and standard deviation are predicted based on a hyper-code. 
See Appendix~\ref{appendix:arch} for more details about the architecture.

%are equipped with a quantizer, a prior model, and (at deployment) an entropy coder in the bottleneck, based on the approach of~\cite{balleVARIATIONALIMAGECOMPRESSION2018}.
%{\color{red}Point to appendix for arch details. Say that all three AEs use exactly the same architecture (but not weights).}

% The architectures of our I- and P-frame codecs follow $\ssf$. In P-frame implementation though, while $\ssf$ uses Gaussian filters to generate a blur stack and uses scale dimension of optical flow to \underline{non-linearly} point to the blur stack, we generate the blur stack using a Gaussian pyramid followed by bilinearly upsampling all pyramid scales to the original resolution and use optical flow scale to \underline{linearly} point to the blur stack.

\subsection{GoP structure}
\label{sec:gop_structure}

\subsubsection{Frame type selection}

I-frame is the least efficient frame type in terms of coding efficiency, next is P-frame, and finally B-frame delivers the best performance. Since $\bepic$ supports all three frame types, it is important to use the right frame type to improve the overall coding efficiency.
%\subsubsection{Reference frames}

In neural video coding, reference frames are often inserted at GoP boundaries. A common practice is to code reference frames as intra and use them as references to code the other frames as inter. In this work, we call this configuration $\ibi$ where GoP boundaries are coded as I-frames and the middle frames are coded as B-frames. See Fig.~\ref{fig:gop_struct}.a for an illustration. On the other hand, given that I-frames are the least efficient among the three frame types, we can code some references as P-frames to improve the performance. Here, we call this configuration $\ibp$ as shown in Fig.~\ref{fig:gop_struct}.b where the first reference is coded as an I-frame and the subsequent references are coded as P-frames~\cite{gop_structure}
\footnotetext{What we show as GoP in this figure is related to a structure of pictures (SoP), but we use GoP liberally because it is more intuitive~\cite{hevcbook}}.

\subsubsection{B-frames order}
Once a B-frame is coded, it can be used as a reference for the next B-frames. It is thus very important to code the B-frames in a GoP in the optimal order to \emph{i)} maximally exploit the information available through the available references and \emph{ii)} derive good references for the next B-frames. 

\begin{figure}[t]
%   \vspace{-.7em}
  \begin{center}
    \includegraphics[width=.8\linewidth]{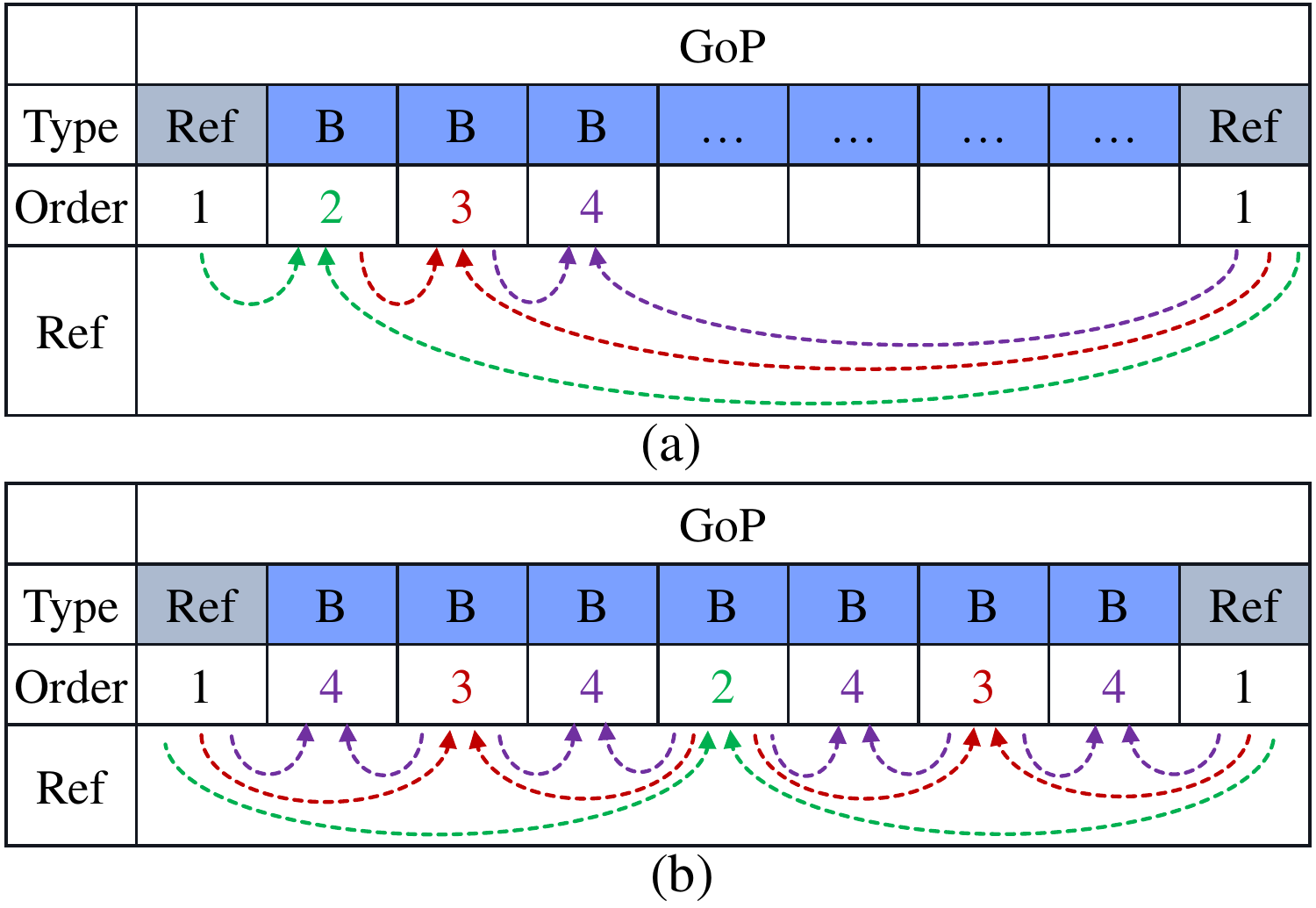}
  \end{center}
  \vspace{-1em}
  \caption{B-frames order, (a) $\seq$, (b) $\hier$.}
  \label{fig:coding_order}
  \vspace{-1em}
\end{figure}

Here, we present two ways to traverse the B-frames in a GoP, $\seq$ and $\hier$~\cite{gop_structure} as shown in Fig.~\ref{fig:coding_order}. We assume that for each given GoP, the boundary frames are references that are already available (decoded) and the middle frames are B-frames.

%For each given GoP, we assume that the boundary frames are references and the middle frames are B-frame. Once a B-frame is coded, it can be used as a reference for the next B-frames. Here, we present two ways to traverse the B-frames in a GoP, $\seq$ and $\hier$ as shown in Fig.~\ref{fig:coding_order}. 

In the $\seq$ order, we start from one end of GoP and code one B-frame at a time until we reach the other end while in the $\hier$ order, we always code the middle frame of two references as the next B-frame. The plain $\hier$ order only supports GoP sizes that are a power of 2 plus 1 \eg 9, 17, 33. However, we devised an algorithm based on bisection to traverse a GoP with an arbitrary size in the $\hier$ order as shown in Algorithm~\ref{alg:hier}.

\begin{algorithm}[!b]
  \caption{Traversing an arbitrary size GoP in $\hier$ order.}
  \label{alg:hier}
  \small
  \SetAlgoLined
    \textbf{Assumption:} Given a GoP of size $N$, suppose $\widehat{\mathbf{x}}_0$ and $\widehat{\mathbf{x}}_{N-1}$ are available as references. \;
    Initialize an empty $\mathbf{stack}$ \;
    \texttt{push} $(i_{\textup{min}}, i_{\textup{max}}) = (0, N-1)$ to $\mathbf{stack}$ \;
    \While {$\mathbf{stack}$ is not empty}{
        \texttt{pop} $\mathbf{stack}$ to retrieve $(i_{\textup{min}}, i_{\textup{max}})$ \;
        Define new index $i = \operatorname{floor}( (i_{\textup{min}} + i_{\textup{max}}) / 2)$ \;
        Encode $\mathbf{x}_i$ as a B-frame to generate $\widehat{\mathbf{x}}_i$. Use $\widehat{\mathbf{x}}_{i_{\textup{min}}}$ and $\widehat{\mathbf{x}}_{i_{\textup{max}}}$ as references \;
        \If {$i - i_{\textup{min}} > 1$} { \texttt{push} $(i_{\textup{min}}, i)$ to $\mathbf{stack}$}
        \If {$i_{\textup{max}} - i > 1$} { \texttt{push} $(i, i_{\textup{max}})$ to $\mathbf{stack}$}
    }
\end{algorithm}

\subsection{Loss function}
%For a video clip consisting of $N$ frames where the first frame is an I-frame, the last frame is a P-frame that uses the I-frame as reference, and the middle frames are B-frames, the loss function is a rate-distortion trade-off as follows:
%{\color{red} Is this true even for IBI mode? Or do we always train like IBP and eval also for IBI? We should mention it here.}

We assume that the training is done on GoPs of $N$ frames whose boundaries are references. The loss function for the $\ibp$ configuration is defined as a rate-distortion tradeoff as follows:

\begin{align}
  \label{eq:loss_ibp}
    \sum_{t=0}^{N-1} D\big(\xt, \: \xhatt \big) + \beta \bigg[ H\big(\textbf{z}_0^i) + \sum_{t=1}^{N-1}\bigg( H\big( \textbf{z}_t^f \big) + H\big( \textbf{z}_t^r \big) \bigg) \bigg]
\end{align}

\noindent where $H(\cdot)$ represents the entropy estimate of a latent in terms of bits-per-pixel which corresponds to the expected size in the bitstream, $\textbf{z}_0^i$ denotes the latent of the I-frame codec's $\iframeae$, $\textbf{z}_t^f$ and $\textbf{z}_t^r$ represent the latents of the P-frame codec's $\flowae$ and $\residualae$, $D(\cdot,\cdot)$ denotes distortion in terms of $\mse$ or $\msssim$~\cite{wangMultiScaleStructuralSimilarity2003}, and $\beta$ is a hyperparameter that controls the balance between rate and distortion. The loss function for $\ibi$ is similar.
%{\color{red} Cite MSSSIM}

\begin{figure*}[t]
  \begin{center}
    \includegraphics[width=.92\textwidth]{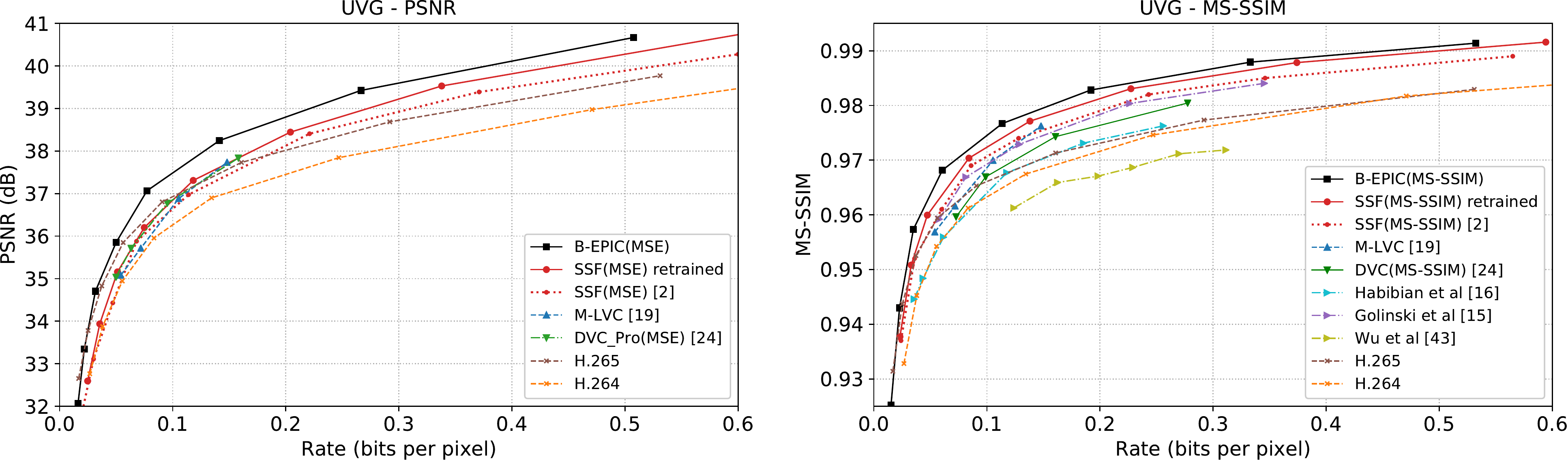}
  \end{center}
  \vspace{-1em}
  \caption{Rate-distortion comparison on the UVG dataset.}
  \label{fig:uvg_rd}
\end{figure*}

\begin{figure*}[tb]
  \begin{center}
    \includegraphics[width=.92\textwidth]{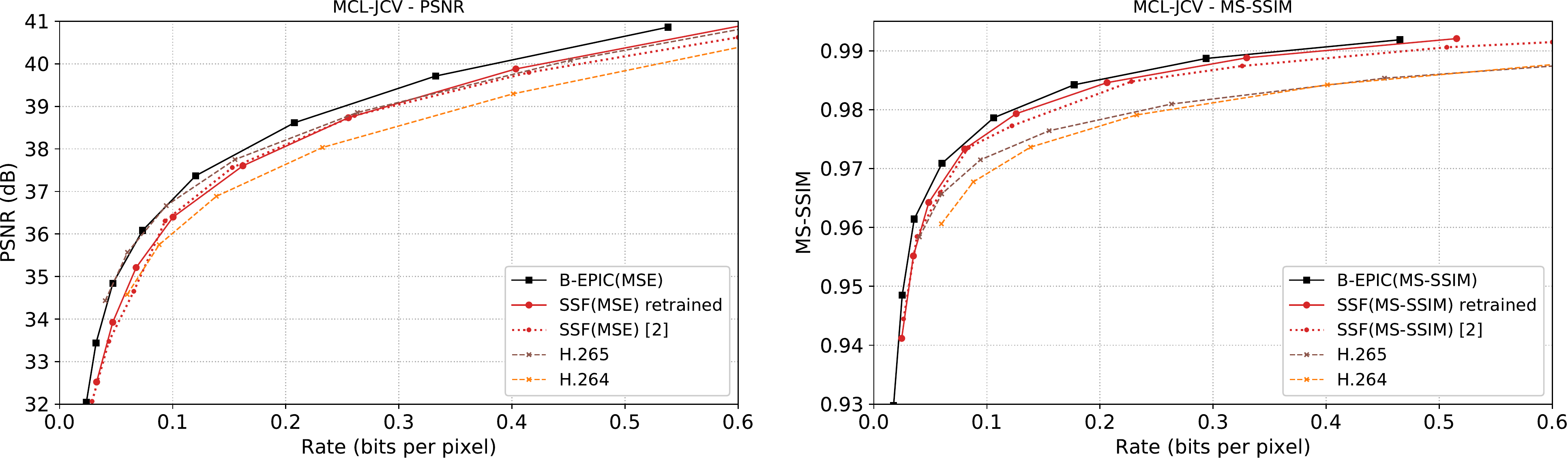}
  \end{center}
  \vspace{-1em}
  \caption{Rate-distortion comparison on the MCL-JCV dataset.}
  \label{fig:mcljcv_rd}
\end{figure*}

\section{Experiments \& Results}
\subsection{Training setup}

\textbf{Dataset:} We used Vimeo-90k as training dataset~\cite{Xue_2019}.
It consists of $89,800$ 7-frame sequences in RGB format.

\textbf{Trained models:} 
We trained models at various rate-distortion tradeoffs ($\beta$ values), using both $\mse$ and $\msssim$ as distortion losses.
The $\mse$ models $\bepicmse$ were trained on $\beta=2^\gamma \times 10^{-4}: \gamma \in \big\{0, 1, ..., 7\big\}$ and the $\msssim$ models $\bepicmsssim$ were trained on $\beta=2^\gamma \times 10^{-2}: \gamma \in \big\{0, 1, ..., 7\big\}$, where both $\mse$ and $\msssim$ were measured in the RGB color space.

%We trained two sets of models, 8 $\mse$ models that deliver good $\psnr$ results and 8 $\msssim$ models that deliver good $\msssim$ results, where each set spans a large range of bit-rates controlled by $\beta$. The $\mse$ models were trained on $\beta=2^\gamma \times 10^{-4}: \gamma \in \big\{0, 1, ..., 7\big\}$ and the $\msssim$ models were trained on $\beta=2^\gamma \times 10^{-2}: \gamma \in \big\{0, 1, ..., 7\big\}$.

\textbf{Training plan:} we followed the exact schedule provided in $\ssf$ \cite{Agustsson_2020_CVPR} to train the $\mse$ and $\msssim$ models to facilitate comparison.
Specifically, we initially trained all models for 1,000,000 gradient updates on $\mse$, then trained the $\msssim$ models for extra 200,000 gradient updates on $\msssim$, and finally fine-tuned all the models for 50,000 gradient updates.

In both training and fine-tuning steps, we employed the Adam optimizer~\cite{kingmaAdamMethodStochastic2015}, used batch size of 8, and trained the network on 4-frame sequences, so the training GoP structure was \texttt{IBBI} and \texttt{IBBP} for the $\ibi$ and $\ibp$ models, respectively. The training step took about 10 days on an Nvidia V100 GPU. We tried other training GoP lengths as well. Longer GoP lengths, despite dramatically slowing down the training, did not improve the performance significantly. In the training step, we set the learning-rate to $10^{-4}$ and used randomly extracted $256\times256$ patches. In the fine-tuning step, we reduced the learning-rate to $10^{-5}$ and increased the patch size to $256\times384$. %{\color{red} Why the change in patch size? Can we give a rationale? If not, we can say ``For the finetuning phase, we decided to use patches of size ...''}

We started all the components in our network from random weights except $\flownet$ in the interpolation block where a pretrained $\operatorname{PWC-Net}$ with frozen weights was employed. 
% {\color{red} How was flownet initialized? From pretraining on ...? Something released by the authors?} 
The gradient from B-frame codecs was stopped from propagating to the I-frame and P-frame codecs for more stable training~\cite{Agustsson_2020_CVPR}.

\subsection{Evaluation setup}
We evaluated $\bepic$ on the UVG~\cite{UVG}, MCL-JCV~\cite{mcljcv}, and HEVC~\cite{HEVC_dataset} datasets, all of which are widely used in the neural video codec literature. All these datasets are available in YUV420. We used $\ffmpeg$~\cite{ffmpeg} to convert the videos to RGB that is acceptable by $\bepic$ (and almost all other neural codecs). See Appendix~\ref{appendix:ffmpeg} for the $\ffmpeg$ commands details.

As shown in section~\ref{sec:ablations}, the $\hier$ B-frame order together with the $\ibp$ GoP structure generate our best results. The results we report in the rest of this section are based on these settings as well as an evaluation GoP size of 12 for consistency with other works~\cite{DVC_TPAMI,wuVideoCompressionImage2018}.
% {\color{red} \cite{}}.

We report video quality in terms of $\psnr$ and $\msssim$ where both are first calculated per-frame in the RGB color space, then averaged over all the frames of each video, and finally averaged over all the videos of a dataset. Due to the network architecture, our codec accepts inputs whose spatial dimensions are a multiple of 64. Whenever there is a size incompatibility, we pad the frame to the nearest multiple of 64 before feeding to the encoder, and crop the decoded frame to compensate for padding.
This issue, which may be fixable, could lead to a coding inefficiency depending on the number of pixels that have to be padded.

\begin{figure*}[tb]
  \begin{center}
    \includegraphics[width=1.0\textwidth]{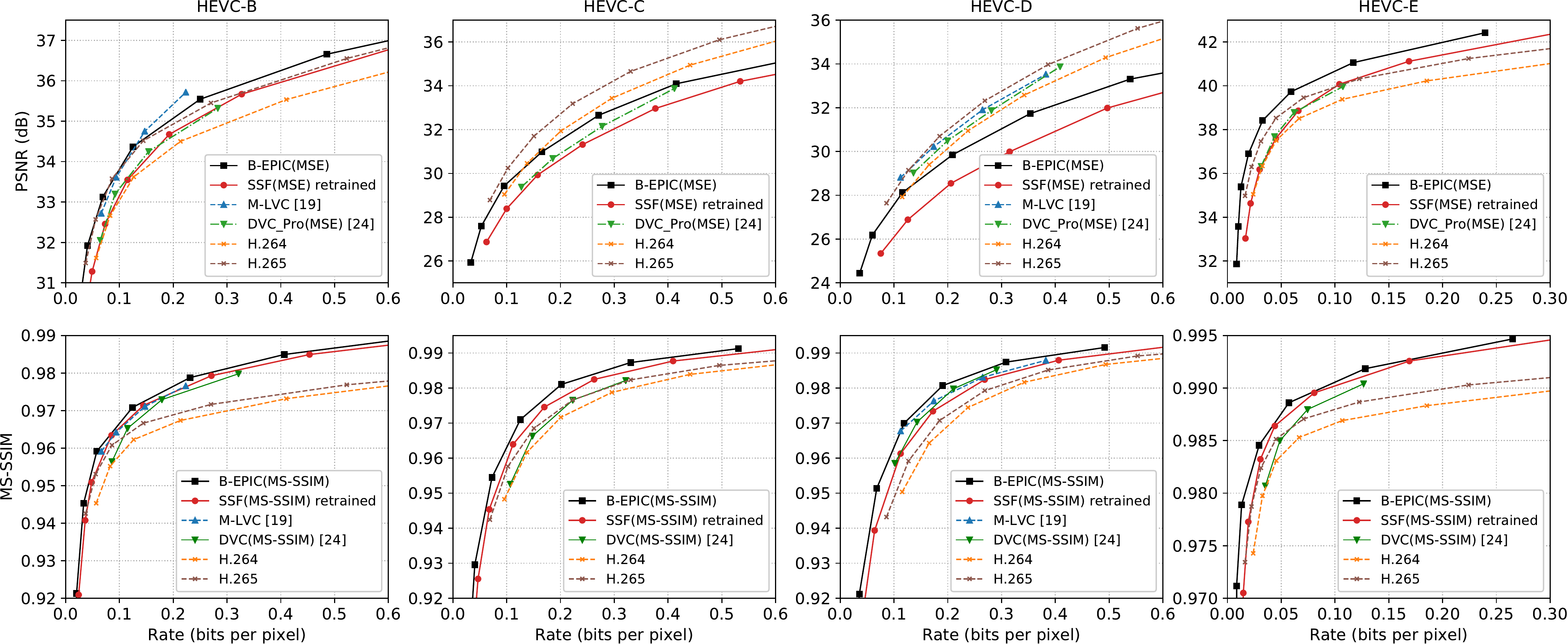}
  \end{center}
  \vspace{-1em}
  \caption{Rate-distortion comparison on the HEVC dataset.}
  \label{fig:hevc_rd}
\end{figure*}

\begin{figure*}[tb]
  \begin{center}
    \includegraphics[width=1.0\textwidth]{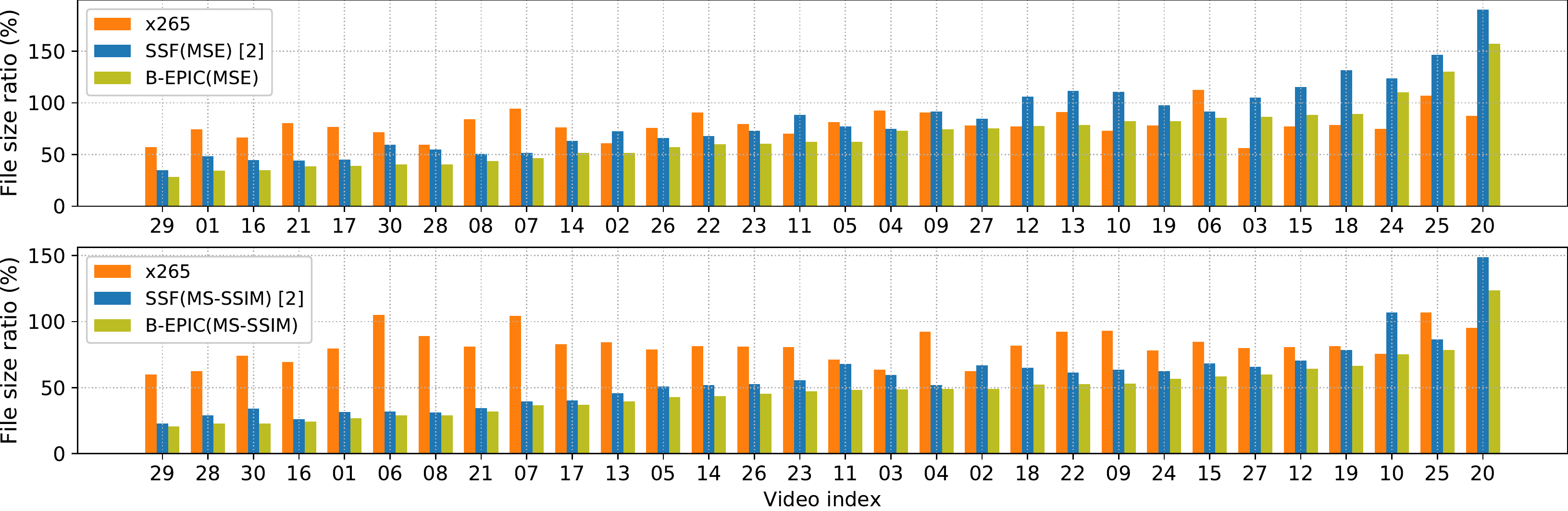}
  \end{center}
  \vspace{-1em}
  \caption{File sizes generated by $\hevc$, $\ssf$, and $\bepic$ relative to $\avc$ on the MCL-JCV dataset, measured by $\psnr$ (top) and $\msssim$ (bottom) BD-rate gain. A value of 100\% translates to equal sizes and no bit-rate savings. Smaller values are preferable.}
  \label{fig:mcljcv_rate}
\end{figure*}

\subsection{Compared methods}
We compared our results with several neural video codecs including $\ssf$~\cite{Agustsson_2020_CVPR}, DVC~\cite{DVC_TPAMI}, M-LVC~\cite{mlvc}, Golinski \etal~\cite{Golinski_2020_ACCV}, Wu \etal~\cite{wuVideoCompressionImage2018}, and Habibian \etal~\cite{Habibian_2019_ICCV}. 
% It is worth mentioning that we
We reimplemented and trained $\ssf$, and here provide both the original results reported in the paper as well as the reproduced results. The results in the original paper where obtained with GoP size of infinity \ie only the first frame of a sequence is an I-frame and the rest are all P-frames while we report the performance on GoP of 12.

The standard codecs that we compare with are $\avc$~\cite{avc} and $\hevc$~\cite{sullivanOverviewHighEfficiency2012}. We generated the results for both using $\ffmpeg$ where GoP size of 12 was used with all the other default settings. This is unlike the other papers that limit the $\ffmpeg$ performance by not allowing B-frames, or changing the preset to \texttt{fast} or \texttt{superfast}. See Appendix~\ref{appendix:ffmpeg} for the $\ffmpeg$ commands details.

\subsection{Results \& Discussion}

\textbf{Rate-distortion}: 
the rate-distortion comparisons on the UVG, MCL-JCV, and HEVC datasets are shown in Figs.~\ref{fig:uvg_rd}, \ref{fig:mcljcv_rd}, \ref{fig:hevc_rd}, respectively. 

When evaluated in terms of $\msssim$, $\bepicmsssim$ outperforms all the compared methods on all the datasets across all bit-rates. 

When evaluated in terms of $\psnr$, as can be observed from Figs.~\ref{fig:uvg_rd} and \ref{fig:mcljcv_rd}, $\bepicmse$ significantly outperforms all the competing neural codecs as well as $\avc$ across all bit-rates on both UVG and MCL-JCV datasets. Compared to $\hevc$, $\bepicmse$ maintains a large margin in the average and high bit-rates and is roughly on-par in extremely low bit-rate cases. On the HEVC dataset, 
% As can be seen from Fig.~\ref{fig:hevc_rd}, our method outperforms all competing methods across all classes and at all bitrates in terms of MS-SSIM.
% When evaluated in terms of PSNR, 
the results are similarly favorable on HEVC class-B and class-E.
On class-C, the standard codecs outperform all neural methods, and on class-D $\bepicmse$ performs poorly.
%On the HEVC dataset, as can be seen from Fig.~\ref{fig:hevc_rd}, our codec behaves the same as UVG and MCL-JCV on class-B and class-E, outperform the compared neural codecs on class-C, but does not perform very well on class-D. 
This is most likely due to the fact that the class-D videos have to be padded from $240 \times 416$ to the nearest multiple of $64$, i.e. $256 \times 448$, before it can be fed to our encoder.
%the size incompatibility of the videos in this class with our network \ie since class-D videos are $240\times416$ they need to be padded to $256\times448$ to make the dimension a factor of 64.
%That is equivalent of encoding extra 15\% pixels that are thrown away in the decoder.
This means our method in its current form has to encode $15\%$ more pixels, all of which are discarded on the decoder side. It is worth noting that HEVC class-D is already removed in the common test conditions of the most recent standard video codec $\operatorname{H.266}$~\cite{vvc10} due to very small resolution.

% It is worth noting that o
$\bepic$ can be thought of as a B-frame equivalent of $\ssf$ and as can be observed from the rate-distortion comparisons, outperforms $\ssf$ significantly across all bit-rates on all the datasets. This proves the effectiveness of our B-frame approach when applied to an existing P-frame codec.

\textbf{Bjøntegaard delta rate (BD-rate)}: 
\label{sec:bd_results}
in this section, we report BD-rate~\cite{bdrate} gains versus $\avc$. Table~\ref{tab:bd_rate} lists the average BD-rate gains versus $\avc$ on the UVG, MCL-JCV, and HEVC datasets in terms of both $\psnr$ and $\msssim$. Here, the numbers show how much a method can save on bit-rate compared to $\avc$ while generating the same video quality. $\bepic$ yields the highest $\msssim$ BD-rate gains and performs relatively well in terms of $\psnr$ BD-rate gains.

\begin{table}[h!]
    \centering
    \scriptsize
    \begin{tabular}{@{\hspace{2pt}}c@{\hspace{3pt}}|@{\hspace{3pt}}c@{\hspace{2pt}}c@{\hspace{2pt}}c@{\hspace{3pt}}|@{\hspace{3pt}}c@{\hspace{2pt}}c@{\hspace{2pt}}c@{\hspace{2pt}}}
        \hline
        & \multicolumn{3}{c}{$\psnr$ BD-rate gain (\%)} & \multicolumn{3}{c}{$\msssim$ BD-rate gain (\%)} \\
        \hline 
        Dataset & $\hevc$ & $\ssf$ & $\bepic$ & $\hevc$ & $\ssf$ & $\bepic$ \\
         &  & $\operatorname{(MSE)}$ & $\operatorname{(MSE)}$ &  & $\operatorname{(MS-SSIM)}$ & $\operatorname{(MS-SSIM)}$ \\
        \hline 
        UVG & -28.66 & -25.71 & \textbf{-47.89} & -24.03 & -41.05 & \textbf{-50.79} \\
        \hline 
        MCL-JCV & -20.86 & -15.90 & \textbf{-31.91} & -18.22 & -43.25 & \textbf{-52.12} \\
        \hline
        HEVC-B & -25.0 & -16.28 & \textbf{-35.68} & -19.70 & -48.23 & \textbf{-55.19} \\
        HEVC-C & \textbf{-19.51} & 47.69 & 11.22 & -15.80 & -27.32 & \textbf{-41.53} \\
        HEVC-D & \textbf{-16.25} & 82.89 & 29.05 & -11.86 & -20.79 & \textbf{-43.98} \\
        HEVC-E & -32.53 & -24.35 & \textbf{-53.87} & -30.45 & -47.81 & \textbf{-61.32} \\
        \hline
        HEVC-Avg & \textbf{-23.32} & 22.49 & -12.32 & -19.46 & -36.04 & \textbf{-50.50} \\
        \hline
    \end{tabular}
    \caption{Average BD-rate gain versus $\avc$ on different datasets.}
    \label{tab:bd_rate}
\end{table}

\begin{figure*}[tb]
  \begin{center}
    \includegraphics[width=1.0\textwidth]{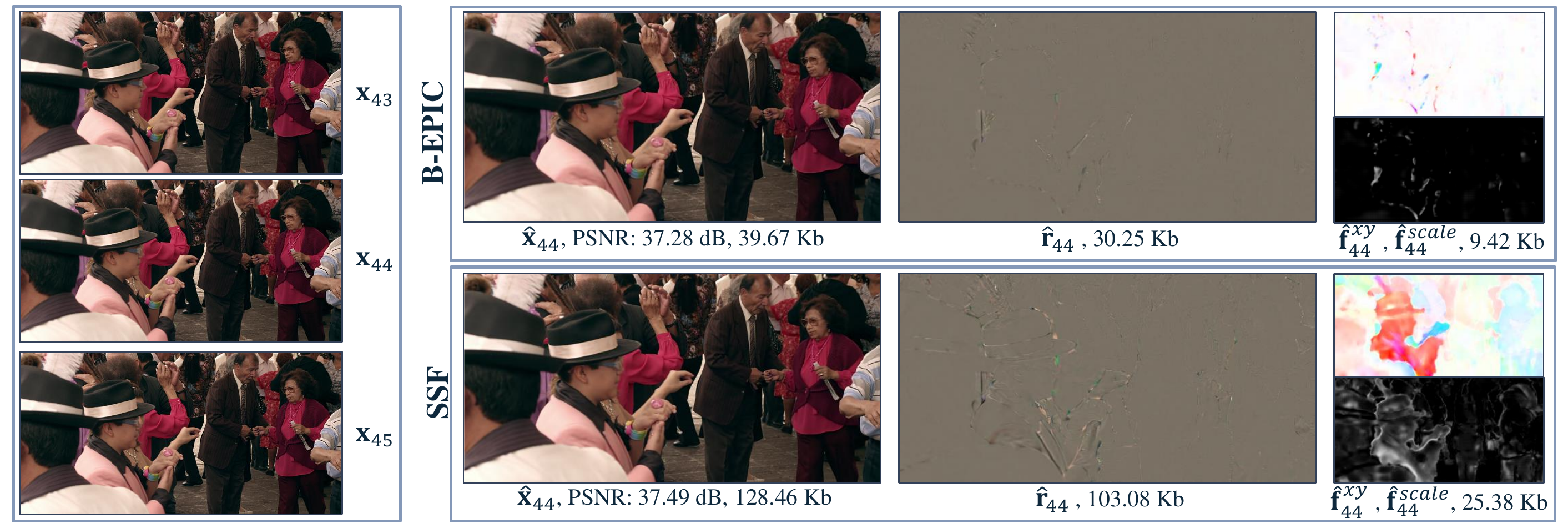}
  \end{center}
  \vspace{-1em}
  \caption[Caption for LOF]{Qualitative results on frame 44 of Tango video from Netflix Tango in Netflix El Fuente\footnotemark~\cite{Xiph}. $\mathbf{x}_{43}$, $\mathbf{x}_{44}$, and $\mathbf{x}_{45}$ are an input sequence. In $\ssf$, $\mathbf{x}_{44}$ is coded as a P-frame with $\mathbf{\widehat{x}}_{43}$ used as reference. In $\bepic$, $\mathbf{x}_{44}$ is coded as a B-frame with both $\mathbf{\widehat{x}}_{43}$ and $\mathbf{\widehat{x}}_{45}$ used as references. The interpolation block delivers an accurate baseline frame used in the P-frame codec as reference. As a result, both flow and residual are less detailed and consume fewer bits compared to $\ssf$.}
  \label{fig:visual}
\end{figure*}

Furthermore, in Fig~\ref{fig:mcljcv_rate} we show the file sizes of the individual MCL-JCV videos encoded using $\bepic$, $\ssf$, and $\hevc$ compared to $\avc$, estimated by BD-rate. As observed here, $\bepic$ delivers better results compared to $\ssf$ across the board. Compared to $\hevc$, it performs significantly better on the majority of the videos, specially in terms of $\msssim$. The under-performance on the last sequences of the figure is potentially because they are animated movies, while our training dataset Vimeo-90k is only comprised of natural videos, as pointed out in \cite{Agustsson_2020_CVPR} as well.
%they are animations while there are not animation in our training dataset, as pointed out in \cite{Agustsson_2020_CVPR} as well.

\textbf{Qualitative results}: 
Fig.~\ref{fig:visual} shows a sample qualitative result where an input sequence together with the decoded frames, optical flow maps, and residuals for $\ssf$ and $\bepic$ are visualized. $\bepic$ relies on much less detailed optical flow and residuals due to the frame-interpolation outcome being used as reference, and as a result, generates a lot less bits compared to $\ssf$ at a similar $\psnr$.

\textbf{Per-frame performance}: Fig.~\ref{fig:uvg_across_gop} shows how $\bepic$ performs on average across a GoP of 12 frames when using the $\seq$ and $\hier$ B-frames orders on the UVG dataset. As expected, as the gap between the two references closes (by newly coded B-frames used as reference), $\psnr$ improves and Rate drops.

\begin{figure}[tb]
  \begin{center}
    \includegraphics[width=1.0\linewidth]{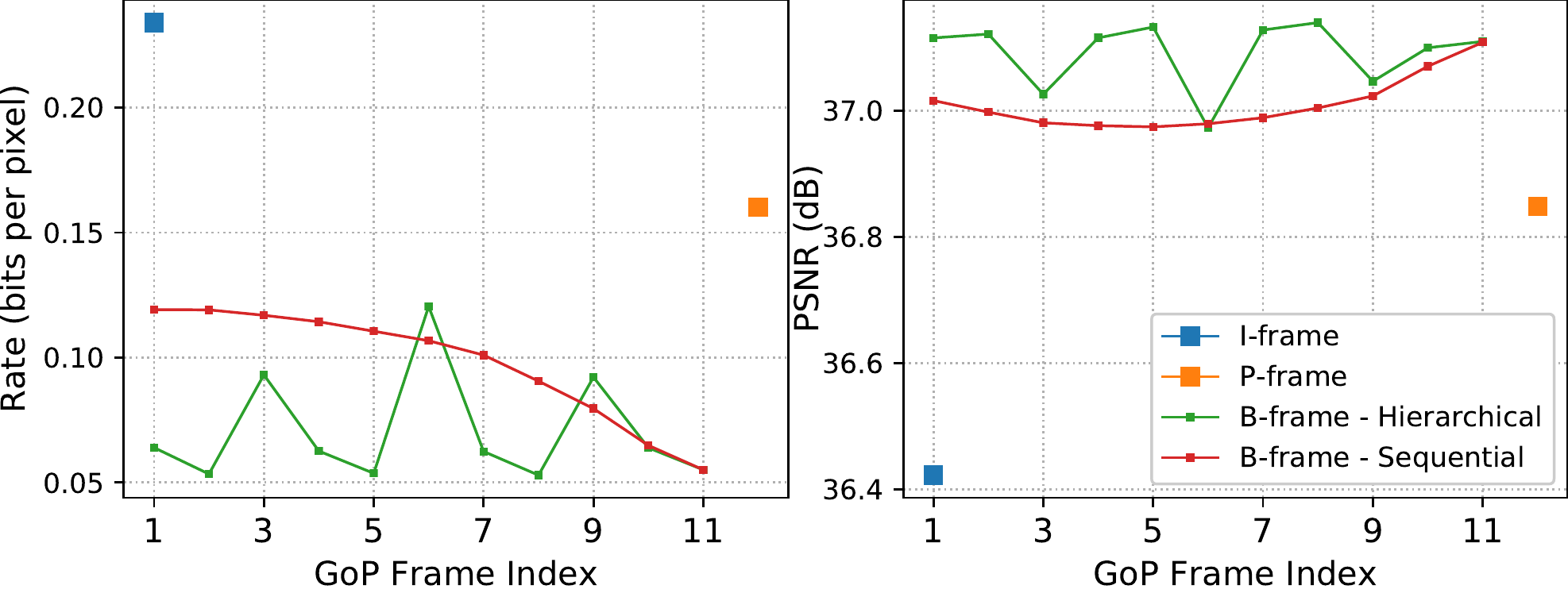}
  \end{center}
  \vspace{-1em}
  \caption{Average per-frame results across a GoP of 12 for the $\seq$ and $\hier$ B-frames orders on the UVG dataset. The first frame of each sequence is coded I, the last frame of each GoP is coded P, and the rest are coded B.}
  \label{fig:uvg_across_gop}
\end{figure}

\subsection{Ablation studies}
\label{sec:ablations}
We studied the effectiveness of different components of our codec including: GoP structure ($\ibi$ vs $\ibp$), B-frames order ($\seq$ vs $\hier$), pretraining $\operatorname{PWC-Net}$, and removing $\flowae$ for the P-frame codec and relying only on $\residualae$. The last configuration where $\flowae$ is removed, is similar to the B-frame codecs that use interpolation followed by residual correction~\cite{Cheng_2019_CVPR}. These ablation studies are shown in Fig.~\ref{fig:ablations}.a. Moreover, we studied the effect of the training GoP on the performance by finetuning our model on different sequences including: 4 consecutive frames, 7 consecutive frames, and 4 frames where consecutive frames are two frames apart. All the studied configurations delivered similar rate-distortion results on the UVG datasets. So, we proceeded with 4 consecutive frames as it is the most memory efficient and fastest to train.
% which is the most convenience and memory efficient one to train our models.

\begin{figure}[tb]
  \begin{center}
    \includegraphics[width=1.0\linewidth]{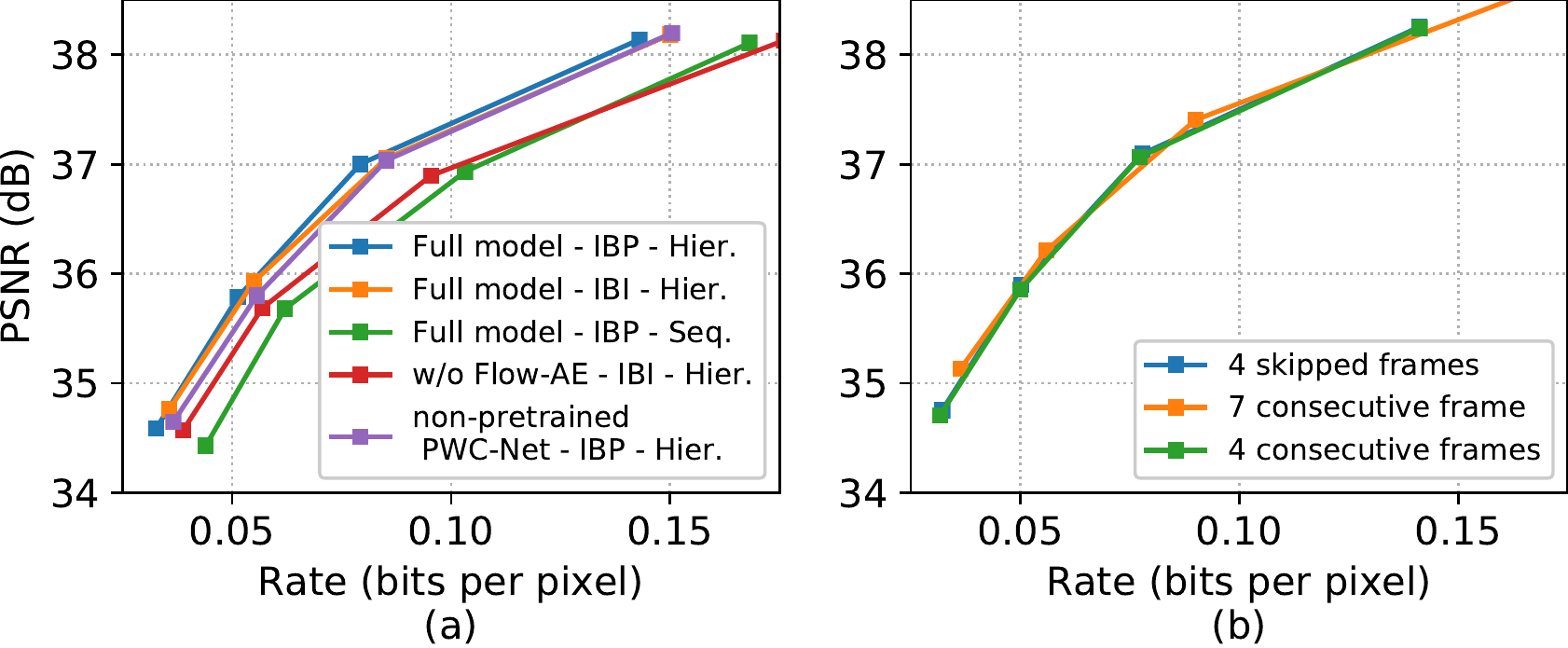}
  \end{center}
  \vspace{-1em}
  \caption{Ablations studies on the UVG dataset, (a) effectiveness of each component, (b) effect of the training sequence length/shape on the performance.}
  \label{fig:ablations}
\end{figure}
\section{Conclusion}
In this paper, we proposed a method to add B-frame coding capability to an existing neural P-frame codec by adding an interpolation block. It significantly improves the performance of the P-frame codec and delivers state-of-the-art neural video coding results on multiple datasets. Since the prototype we developed in this work is based on 2-reference B-frames and 1-reference P-frame codecs, as a future direction, this idea can be extended to the cases where more that 2 references are available to B-frames and/or with multi-frame P-frame codecs
\footnotetext{
  \tiny{Video produced by Netflix, with \texttt{CC BY-NC-ND 4.0} license: \\
  \url{https://media.xiph.org/video/derf/ElFuente/Netflix_Tango_Copyright.txt}}}

\clearpage
%%%%%%%%% Bibliography

{\small
\bibliographystyle{ieee_fullname}
\bibliography{references}

\begin{thebibliography}{10}\itemsep=-1pt

\bibitem{agustssonSofttoHardVectorQuantization2017}
Eirikur Agustsson, Fabian Mentzer, Michael Tschannen, Lukas Cavigelli, Radu
  Timofte, Luca Benini, and Luc~V Gool.
\newblock Soft-to-hard vector quantization for end-to-end learning compressible
  representations.
\newblock In {\em Advances in Neural Information Processing Systems},
  volume~30. Curran Associates, Inc., 2017.

\bibitem{Agustsson_2020_CVPR}
Eirikur Agustsson, David Minnen, Nick Johnston, Johannes Balle, Sung~Jin Hwang,
  and George Toderici.
\newblock Scale-space flow for end-to-end optimized video compression.
\newblock In {\em Proceedings of the IEEE/CVF Conference on Computer Vision and
  Pattern Recognition (CVPR)}, June 2020.

\bibitem{balleDensity2015}
Johannes Ball{\'e}, Valero Laparra, and {Eero P.} Simoncelli.
\newblock Density modeling of images using a generalized normalization
  transformation.
\newblock Jan. 2016.
\newblock 4th International Conference on Learning Representations, ICLR 2016.

\bibitem{balleEndtoendOptimizationNonlinear2016}
J. {Ballé}, V. {Laparra}, and E.~P. {Simoncelli}.
\newblock End-to-end optimization of nonlinear transform codes for perceptual
  quality.
\newblock In {\em 2016 Picture Coding Symposium (PCS)}, pages 1--5, 2016.

\bibitem{balleVARIATIONALIMAGECOMPRESSION2018}
Johannes Ballé, David Minnen, Saurabh Singh, Sung~Jin Hwang, and Nick
  Johnston.
\newblock Variational image compression with a scale hyperprior.
\newblock In {\em International Conference on Learning Representations}, 2018.

\bibitem{bdrate}
Gisle Bjøntegaard.
\newblock Calculation of average {{PSNR}} differences between {{RD-curves}}.
  {{Doc. VCEG-M33}}.
\newblock {ITU-T SG16/Q6 VCEG}, Austin, TX, USA, July 2001.

\bibitem{HEVC_dataset}
Frank Bossen.
\newblock Common test conditions and software reference configurations.
\newblock JCTVC-F900, 2011.

\bibitem{vvc10}
B. Bross, J. Chen, S. Liu, and Y.-K. Wang.
\newblock Versatile video coding (draft 10).
\newblock Output document {JVET-S2001}, July 2020.

\bibitem{chen2020learning}
Meixu Chen, Todd Goodall, Anjul Patney, and Alan~C. Bovik.
\newblock Learning to compress videos without computing motion, 2020.

\bibitem{liuNLAICImageCompression2019}
T. {Chen}, H. {Liu}, Z. {Ma}, Q. {Shen}, X. {Cao}, and Y. {Wang}.
\newblock End-to-end learnt image compression via non-local attention
  optimization and improved context modeling.
\newblock {\em IEEE Transactions on Image Processing}, 30:3179--3191, 2021.

\bibitem{Cheng_2019_CVPR}
Zhengxue Cheng, Heming Sun, Masaru Takeuchi, and Jiro Katto.
\newblock Learning image and video compression through spatial-temporal energy
  compaction.
\newblock In {\em Proceedings of the IEEE/CVF Conference on Computer Vision and
  Pattern Recognition (CVPR)}, June 2019.

\bibitem{Choi_2019_ICCV}
Yoojin Choi, Mostafa El-Khamy, and Jungwon Lee.
\newblock Variable rate deep image compression with a conditional autoencoder.
\newblock In {\em Proceedings of the IEEE/CVF International Conference on
  Computer Vision (ICCV)}, October 2019.

\bibitem{Djelouah_2019_ICCV}
Abdelaziz Djelouah, Joaquim Campos, Simone Schaub-Meyer, and Christopher
  Schroers.
\newblock Neural inter-frame compression for video coding.
\newblock In {\em Proceedings of the IEEE/CVF International Conference on
  Computer Vision (ICCV)}, October 2019.

\bibitem{ffmpeg}
{ffmpeg Developers}.
\newblock ffmpeg.
\newblock \url{http://ffmpeg.org/}.
\newblock Accessed: 2020-02-21.

\bibitem{Golinski_2020_ACCV}
Adam Golinski, Reza Pourreza, Yang Yang, Guillaume Sautiere, and Taco~S. Cohen.
\newblock Feedback recurrent autoencoder for video compression.
\newblock In {\em Proceedings of the Asian Conference on Computer Vision
  (ACCV)}, November 2020.

\bibitem{Habibian_2019_ICCV}
Amirhossein Habibian, Ties~van Rozendaal, Jakub~M. Tomczak, and Taco~S. Cohen.
\newblock Video compression with rate-distortion autoencoders.
\newblock In {\em Proceedings of the IEEE/CVF International Conference on
  Computer Vision (ICCV)}, October 2019.

\bibitem{superslomo}
Huaizu Jiang, Deqing Sun, Varun Jampani, Ming-Hsuan Yang, Erik Learned-Miller,
  and Jan Kautz.
\newblock Super slomo: High quality estimation of multiple intermediate frames
  for video interpolation.
\newblock In {\em Proceedings of the IEEE Conference on Computer Vision and
  Pattern Recognition (CVPR)}, June 2018.

\bibitem{kingmaAdamMethodStochastic2015}
Diederik~P. Kingma and Jimmy Ba.
\newblock Adam: A method for stochastic optimization, 2015.
\newblock International Conference for Learning Representations, San Diego,
  2015.

\bibitem{mlvc}
Jianping Lin, Dong Liu, Houqiang Li, and Feng Wu.
\newblock {M-LVC}: Multiple frames prediction for learned video compression.
\newblock In {\em Proceedings of the IEEE/CVF Conference on Computer Vision and
  Pattern Recognition (CVPR)}, June 2020.

\bibitem{9247134}
H. {Liu}, M. {Lu}, Z. {Ma}, F. {Wang}, Z. {Xie}, X. {Cao}, and Y. {Wang}.
\newblock Neural video coding using multiscale motion compensation and
  spatiotemporal context model.
\newblock {\em IEEE Transactions on Circuits and Systems for Video Technology},
  pages 1--1, 2020.

\bibitem{liu2020learned}
Haojie Liu, Han Shen, Lichao Huang, Ming Lu, Tong Chen, and Zhan Ma.
\newblock Learned video compression via joint spatial-temporal correlation
  exploration.
\newblock {\em Proceedings of the AAAI Conference on Artificial Intelligence},
  34(07):11580--11587, Apr. 2020.

\bibitem{stephan_2019_neurips}
Salvator Lombardo, Jun Han, Christopher Schroers, and Stephan Mandt.
\newblock Deep generative video compression.
\newblock In {\em Advances in Neural Information Processing Systems},
  volume~32. Curran Associates, Inc., 2019.

\bibitem{luDVCEndtoendDeep2019}
Guo Lu, Wanli Ouyang, Dong Xu, Xiaoyun Zhang, Chunlei Cai, and Zhiyong Gao.
\newblock {DVC}: An end-to-end deep video compression framework.
\newblock In {\em Proceedings of the IEEE/CVF Conference on Computer Vision and
  Pattern Recognition (CVPR)}, June 2019.

\bibitem{DVC_TPAMI}
G. {Lu}, X. {Zhang}, W. {Ouyang}, L. {Chen}, Z. {Gao}, and D. {Xu}.
\newblock An end-to-end learning framework for video compression.
\newblock {\em IEEE Transactions on Pattern Analysis and Machine Intelligence},
  2020.

\bibitem{lu2021progressive}
Yadong Lu, Yinhao Zhu, Yang Yang, Amir Said, and Taco~S Cohen.
\newblock Progressive neural image compression with nested quantization and
  latent ordering, 2021.

\bibitem{mentzerConditionalProbabilityModels2018}
Fabian Mentzer, Eirikur Agustsson, Michael Tschannen, Radu Timofte, and Luc
  Van~Gool.
\newblock Conditional probability models for deep image compression.
\newblock In {\em Proceedings of the IEEE Conference on Computer Vision and
  Pattern Recognition (CVPR)}, June 2018.

\bibitem{minnenJointAutoregressiveAndHierarchicalPriors2018}
David Minnen, Johannes Ball\'{e}, and George~D Toderici.
\newblock Joint autoregressive and hierarchical priors for learned image
  compression.
\newblock In S. Bengio, H. Wallach, H. Larochelle, K. Grauman, N. Cesa-Bianchi,
  and R. Garnett, editors, {\em Advances in Neural Information Processing
  Systems}, volume~31. Curran Associates, Inc., 2018.

\bibitem{minnen2020channelwise}
David Minnen and Saurabh Singh.
\newblock Channel-wise autoregressive entropy models for learned image
  compression.
\newblock In {\em IEEE International Conference on Image Processing (ICIP)},
  2020.

\bibitem{9300040}
W. {Park} and M. {Kim}.
\newblock Deep predictive video compression using mode-selective uni- and
  bi-directional predictions based on multi-frame hypothesis.
\newblock {\em IEEE Access}, 9:72--85, 2021.

\bibitem{rippel2017real}
Oren Rippel and Lubomir Bourdev.
\newblock Real-time adaptive image compression.
\newblock In {\em Proceedings of the 34th International Conference on Machine
  Learning}, volume~70, pages 2922--2930, 06--11 Aug 2017.

\bibitem{rippelLearnedVideoCompression2018}
Oren Rippel, Sanjay Nair, Carissa Lew, Steve Branson, Alexander~G. Anderson,
  and Lubomir Bourdev.
\newblock Learned video compression.
\newblock In {\em Proceedings of the IEEE/CVF International Conference on
  Computer Vision (ICCV)}, October 2019.

\bibitem{UNet}
Olaf Ronneberger, Philipp Fischer, and Thomas Brox.
\newblock U-net: Convolutional networks for biomedical image segmentation.
\newblock In {\em Medical Image Computing and Computer-Assisted Intervention --
  MICCAI 2015}, pages 234--241, Cham, 2015. Springer International Publishing.

\bibitem{gop_structure}
H. {Schwarz}, D. {Marpe}, and T. {Wiegand}.
\newblock Analysis of hierarchical b pictures and mctf.
\newblock In {\em 2006 IEEE International Conference on Multimedia and Expo},
  pages 1929--1932, 2006.

\bibitem{sullivanOverviewHighEfficiency2012}
G~J Sullivan, J~R Ohm, W~J Han, and T Wiegand.
\newblock {Overview of the High Efficiency Video Coding (HEVC) Standard}.
\newblock {\em IEEE Trans. Circuits Syst. Video Technol.}, 22(12):1649--1668,
  Dec. 2012.

\bibitem{Sun2018PWC-Net}
D. {Sun}, X. {Yang}, M. {Liu}, and J. {Kautz}.
\newblock {PWC-Net}: Cnns for optical flow using pyramid, warping, and cost
  volume.
\newblock In {\em 2018 IEEE/CVF Conference on Computer Vision and Pattern
  Recognition}, pages 8934--8943, 2018.

\bibitem{todericiVariableRateImage2015}
George Toderici, Sean~M. O'Malley, Sung~Jin Hwang, Damien Vincent, David
  Minnen, Shumeet Baluja, Michele Covell, and Rahul Sukthankar.
\newblock Variable rate image compression with recurrent neural networks.
\newblock {\em CoRR}, abs/1511.06085, 2015.

\bibitem{todericiFullResolutionImage2017}
George Toderici, Damien Vincent, Nick Johnston, Sung Jin~Hwang, David Minnen,
  Joel Shor, and Michele Covell.
\newblock Full resolution image compression with recurrent neural networks.
\newblock In {\em Proceedings of the IEEE Conference on Computer Vision and
  Pattern Recognition (CVPR)}, July 2017.

\bibitem{UVG}
{Ultra Video Group}.
\newblock {UVG} test sequences.
\newblock \url{http://ultravideo.cs.tut.fi/}.
\newblock Accessed: 2020-02-21.

\bibitem{mcljcv}
H. {Wang}, W. {Gan}, S. {Hu}, J.~Y. {Lin}, L. {Jin}, L. {Song}, P. {Wang}, I.
  {Katsavounidis}, A. {Aaron}, and C.~.~J. {Kuo}.
\newblock Mcl-jcv: A jnd-based h.264/avc video quality assessment dataset.
\newblock In {\em 2016 IEEE International Conference on Image Processing
  (ICIP)}, pages 1509--1513, 2016.

\bibitem{wangMultiScaleStructuralSimilarity2003}
Zhou Wang, Alan~C Bovik, Hamid~R Sheikh, Eero~P Simoncelli, et~al.
\newblock {Image quality assessment: from error visibility to structural
  similarity}.
\newblock {\em IEEE Trans. on Image Processing}, 13(4):600--612, 2004.

\bibitem{avc}
T. {Wiegand}, G.~J. {Sullivan}, G. {Bjontegaard}, and A. {Luthra}.
\newblock {Overview of the H.264/AVC video coding standard}.
\newblock {\em IEEE Transactions on Circuits and Systems for Video Technology},
  13(7):560--576, July 2003.

\bibitem{hevcbook}
Mathias Wien.
\newblock {\em High Efficiency Video Coding: Coding Tools and Specification}.
\newblock Springer Publishing Company, Incorporated, 2014.

\bibitem{wuVideoCompressionImage2018}
Chao-Yuan Wu, Nayan Singhal, and Philipp Krahenbuhl.
\newblock Video compression through image interpolation.
\newblock In {\em Proceedings of the European Conference on Computer Vision
  (ECCV)}, September 2018.

\bibitem{Xiph}
Xiph.org.
\newblock Xiph.org video test media [derf's collection].
\newblock \url{https://media.xiph.org/video/derf/}.
\newblock Accessed: 2020-02-21.

\bibitem{quadratic_NEURIPS2019_d045c59a}
Xiangyu Xu, Li Siyao, Wenxiu Sun, Qian Yin, and Ming-Hsuan Yang.
\newblock Quadratic video interpolation.
\newblock In {\em Advances in Neural Information Processing Systems},
  volume~32. Curran Associates, Inc., 2019.

\bibitem{Xue_2019}
Tianfan Xue, Baian Chen, Jiajun Wu, Donglai Wei, and William~T Freeman.
\newblock Video enhancement with task-oriented flow.
\newblock {\em International Journal of Computer Vision (IJCV)},
  127(8):1106--1125, 2019.

\bibitem{yang2021hierarchical}
Ruihan Yang, Yibo Yang, Joseph Marino, and Stephan Mandt.
\newblock Hierarchical autoregressive modeling for neural video compression.
\newblock In {\em International Conference on Learning Representations}, 2021.

\end{thebibliography}
}

\clearpage
%%%%%%%%% appendix
\appendix
% \counterwithin{figure}{section}
% \counterwithin{table}{section}
{\Large\noindent\bf Appendix}
\section{Architecture details}
\label{appendix:arch}

\subsection{I-/P-frame codec}
The details of the main autoencoder and the hyperprior in $\iframeae$, $\flowae$, and $\residualae$ are shown in Figs.~\ref{fig:codec} and \ref{fig:hypercodec}, respectively.

\begin{figure}[h]
  \begin{center}
    \includegraphics[width=.6\linewidth]{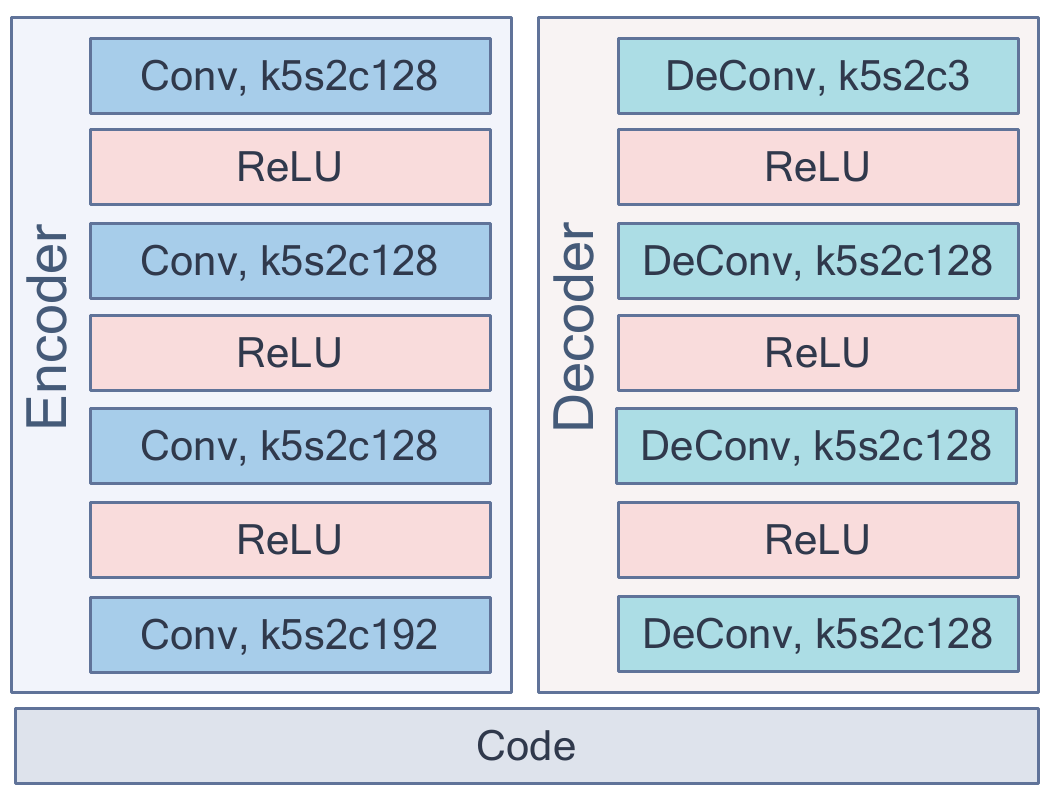}
  \end{center}
  \vspace{-1em}
  \caption{Main autoencoder details. $k$, $s$, and $c$ denote kernel size, stride, and the number of output channels, respectively.}
  \label{fig:codec}
\end{figure}

\begin{figure}[h]
  \begin{center}
    \includegraphics[width=.9\linewidth]{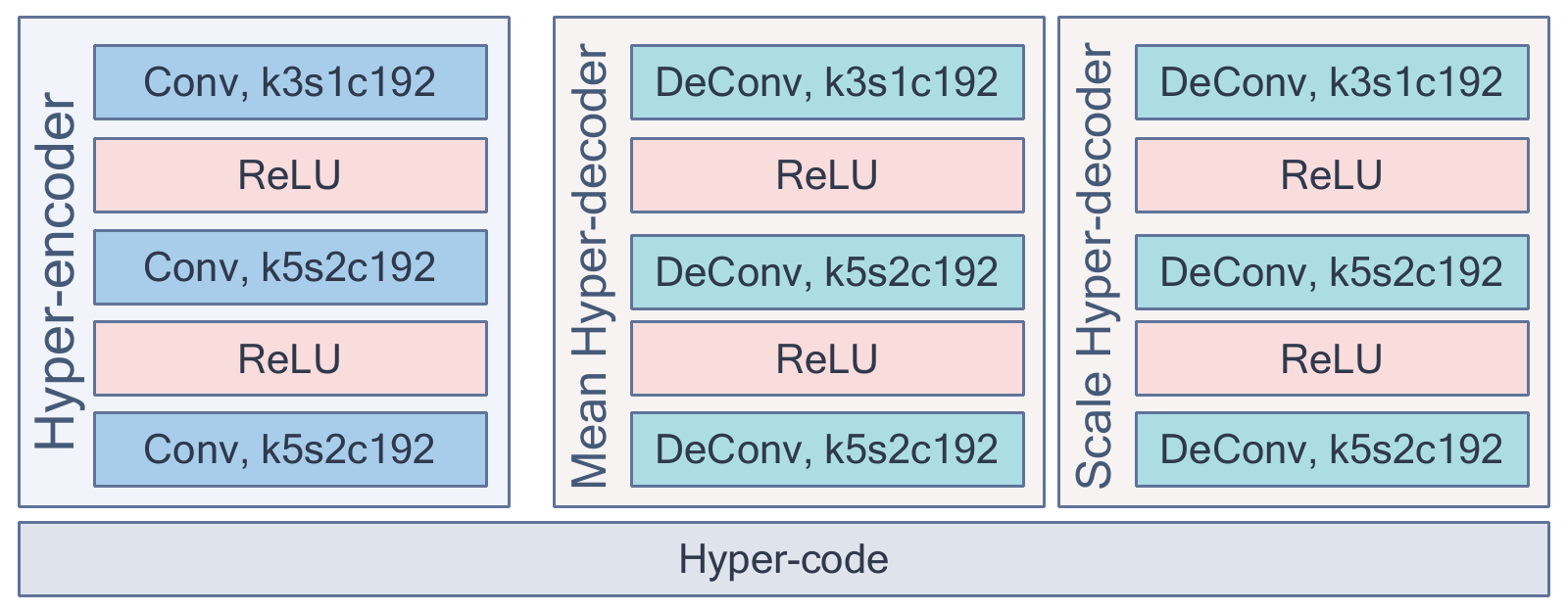}
  \end{center}
  \vspace{-1em}
  \caption{Hyperprior details. $k$, $s$, and $c$ denote kernel size, stride, and the number of output channels, respectively.}
  \label{fig:hypercodec}
\end{figure}

In our implementation of the hyperprior, we followed \cite{balleVARIATIONALIMAGECOMPRESSION2018} which differs slightly from $\ssf$~\cite{Agustsson_2020_CVPR} as pointed below:
\begin{itemize}
    \item in the scale hyper-decoder, $\operatorname{QReLU}$ activations are replaced by $\operatorname{ReLU}$ and the last $\operatorname{QReLU}$ is removed. To have a lower bound on standard deviation values, we clamp the scale hyper-decoder output at $0.11$,
    \item in both hyper-decoders, the last layer is implemented as a $\operatorname{DeConv} \: 3\times3$ with stride 1 as opposed to $\operatorname{DeConv} \: 5\times5$ with stride 2 in $\ssf$,
    \item the hyper-encoder, the first layer is implemented as a $\operatorname{Conv} \: 3\times3$ with stride 1 as opposed to $\operatorname{Conv} \: 5\times5$ with stride 2 in $\ssf$.
\end{itemize}

\subsection{Frame interpolation}

In the frame interpolation component, $\flownet$ is a pre-trained $\operatorname{PWC-Net}$~\cite{Sun2018PWC-Net} without modifications and $\refinenet$ is a $\operatorname{U-Net}$ shown in Fig.~\ref{fig:unet}.

\begin{figure}[h]
  \begin{center}
    \includegraphics[width=.8\linewidth]{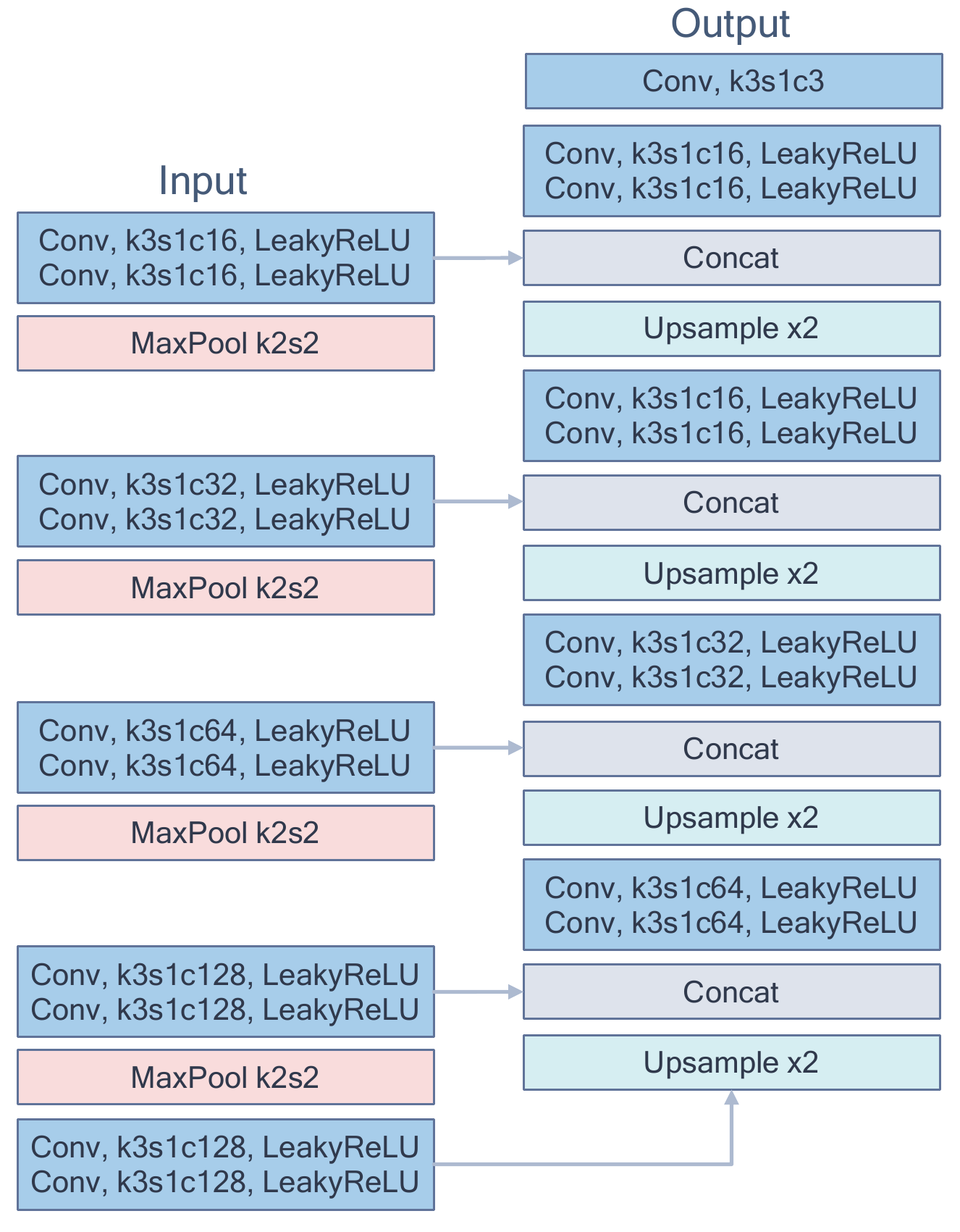}
  \end{center}
  \vspace{-1em}
  \caption{U-Net architecture details. $k$, $s$, and $c$ denote kernel size, stride, and the number of output channels, respectively.}
  \label{fig:unet}
\end{figure}

\section{FFMPEG commands}
\label{appendix:ffmpeg}

We generated $\avc$ and $\hevc$ baselines using $\ffmpeg$. The command that we used to run $\ffmpeg$ with all the default configurations is as follows:

\begin{quote}
\noindent\small \texttt{ffmpeg -pix\textunderscore fmt yuv420p -s [W]x[H] -r [FR] -i [IN].yuv -c:v libx[ENC] -b:v -crf [CRF] [OUT].mkv}
\end{quote}

\noindent and the command that we used to run $\ffmpeg$ with GoP=12 is as follows:

\begin{quote}
\noindent\small \texttt{ffmpeg -pix\textunderscore fmt yuv420p -s [W]x[H] -r [FR] -i [IN].yuv -c:v libx[ENC] -b:v -crf [CRF] -x[ENC]-params "keyint=[GOP]:min-keyint=[GOP]:verbose=1" [OUT].mkv}
\end{quote}

\noindent where the values in brackets represent the encoder parameters as follows: \texttt{H} and \texttt{W} are the frame dimensions, \texttt{FR} is the frame rate, \texttt{ENC} is the encoder type (\texttt{x264} or \texttt{x265}), \texttt{GOP} is the GoP size (12),  \texttt{INPUT} and \texttt{OUTPUT} are the input and the output filenames, respectively, and \texttt{CRF} controls the bit-rate (We tried \texttt{CRF}$=\{9, 12, 15, 18, 21, 24, 27, 30\}$).

In order to measure $\msssim$ and $\psnr$ in the RGB color space, we saved all video frames as PNG files using $\ffmpeg$ using the following commands for YUV and MKV files:

\begin{quote}
\noindent\small  \texttt{ffmpeg -pix\textunderscore fmt yuv420p -s [W]x[H] -i [IN].yuv 
\%9d.png}
\end{quote}

\begin{quote}
\noindent\small  \texttt{ffmpeg [IN].mkv \%9d.png}
\end{quote}
\section{Extended results}

\textbf{Bjøntegaard delta rate (BD-rate) comparison}: 
We report the BD-rate~\cite{bdrate} gains for a scenario where both $\avc$ and $\hevc$
are configured to use all the default parameters as opposed to the results for GoP=12 reported in~\ref{sec:bd_results} (see Table~\ref{tab:bd_rate_default}).

\begin{table}[h]
    \centering
    \scriptsize
    \begin{tabular}{@{\hspace{2pt}}c@{\hspace{3pt}}|@{\hspace{3pt}}c@{\hspace{2pt}}c@{\hspace{2pt}}c@{\hspace{3pt}}|@{\hspace{3pt}}c@{\hspace{2pt}}c@{\hspace{2pt}}c@{\hspace{2pt}}}
        \hline
        & \multicolumn{3}{c}{$\psnr$ BD-rate gain (\%)} & \multicolumn{3}{c}{$\msssim$ BD-rate gain (\%)} \\
        \hline 
        Dataset & $\hevc$ & $\ssf$ & $\bepic$ & $\hevc$ & $\ssf$ & $\bepic$ \\
         &  & $\operatorname{(MSE)}$ & $\operatorname{(MSE)}$ &  & $\operatorname{(MS-SSIM)}$ & $\operatorname{(MS-SSIM)}$ \\
        \hline UVG & \textbf{-29.02} & 11.05 & -26.93 & -24.28 & -11.55 & \textbf{-25.99} \\
        \hline 
        MCL-JCV & \textbf{-20.06} & 3.77 & -17.27 & -16.76 & -29.69 & \textbf{-40.76} \\
        \hline
        HEVC-B & \textbf{-22.44} & 6.69 & -18.21 & -16.54 & -32.36 & \textbf{-41.80} \\
        HEVC-C & \textbf{-12.96} & 91.98 & 43.38 & -8.21 & -5.24 & \textbf{-24.66} \\
        HEVC-D & \textbf{-7.64} & 149.80 & 72.16 & -3.13 & 5.71 & \textbf{-27.71} \\
        HEVC-E & \textbf{-27.31} & 34.28 & -18.12 & -27.24 & -4.36 & \textbf{-30.77} \\
        \hline
        HEVC-Avg & \textbf{-17.59} & 70.69 & 19.80 & -13.78 & -9.05 & \textbf{-31.23} \\
        \hline
    \end{tabular}
    \caption{Average BD-rate gain versus $\avc$ (with $\ffmpeg$ default parameters) on different datasets.}
    \label{tab:bd_rate_default}
\end{table}

\textbf{Qualitative results}: 

In Fig.~\ref{fig:vis_gop} we show the intermediate visualizations as well as the detailed rate-distortion results across a GoP of seven frames for both $\bepic$ and $\ssf$. Figures~\ref{fig:vis_perc_1} and \ref{fig:vis_perc_2} show qualitative comparisons of $\bepic$ and $\ssf$ for the first GoP of two videos.

\begin{figure*}[t]
  \begin{center}
    \includegraphics[width=1.0\textwidth]{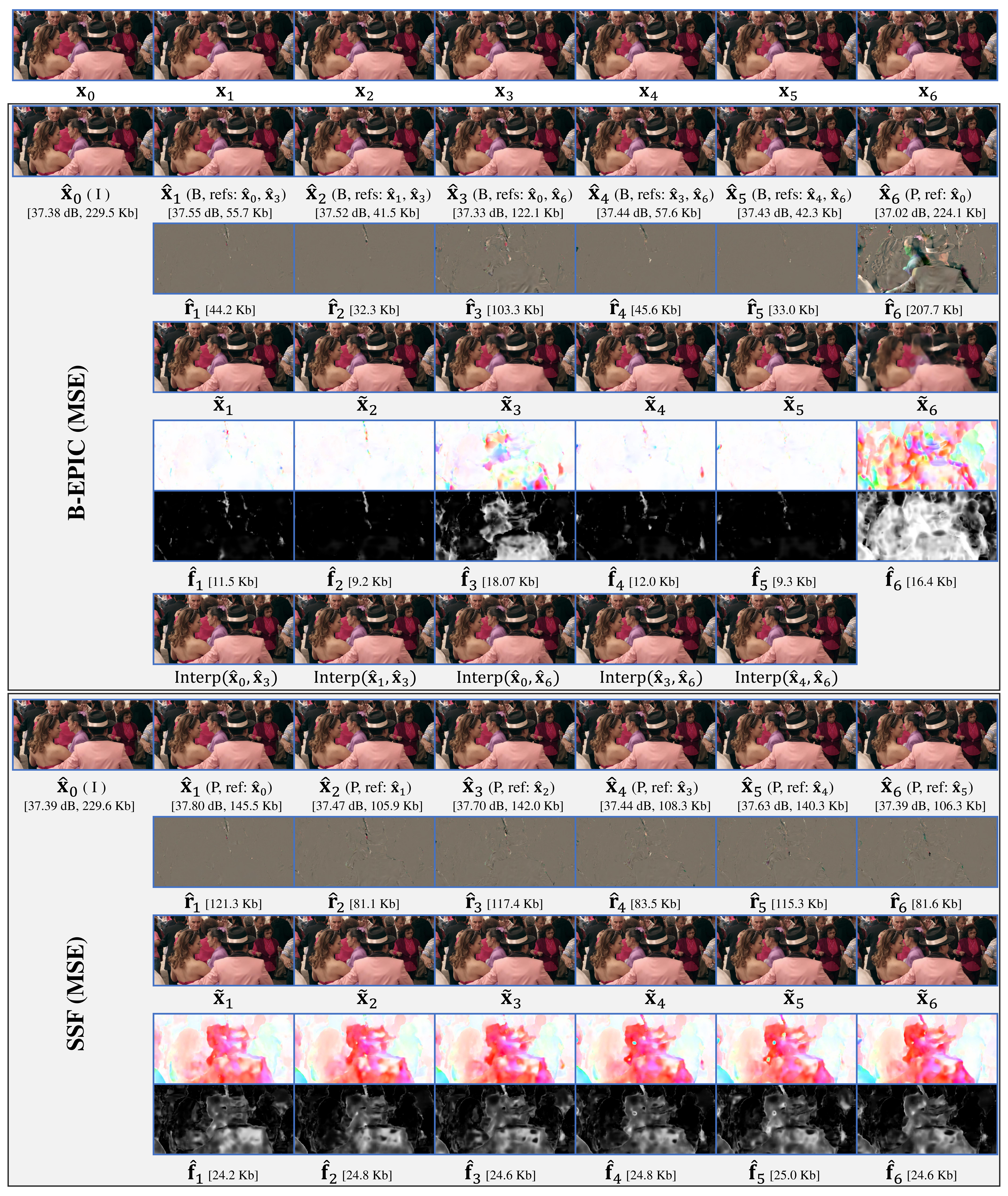}
  \end{center}
  \vspace{-1em}
  \caption{Qualitative results for the first GoP (GoP=7) of Tango video from Netflix Tango in Netflix El Fuente~\cite{Xiph} (resolution $1024 \times 2048$, zoom in for more details). Both $\bepic$ and $\ssf$ models are trained on $\mse$ with $\beta=0.0016$. $\mathbf{x}_0$ is an intra frame with similar performance on $\bepic$ and $\ssf$. The other six frames are inter frames (P or B). $\ssf$ yields consistent bit-rates across the inter frames due to similar level of details in the optical flow and the residuals. That is mainly because all the inter frames are P-frames where the immediate previous decoded frame is used as reference. In $\bepic$, the inter frames are mostly B-frames where the distances to the references are quite variable. As a result, it delivers different bit-rates across the inter frames. The average results for the inter-frames for $\bepic$ and $\ssf$ are as follows: $\psnr$: 37.37dB vs 37.57dB, bit-rate: 90.5 Kb vs 124.7 kb, residual bit-rate: 77.7 Kb vs 100.0 Kb, flow bit-rate: 12.8 Kb vs 24.7 Kb. Here, the bit-rate values are obtained by multiplying $\rate$ (bits-per-pixel) by frame resolution.\\
  \scriptsize{[Video produced by Netflix, with \texttt{CC BY-NC-ND 4.0} license: 
  \url{https://media.xiph.org/video/derf/ElFuente/Netflix_Tango_Copyright.txt}}]}
  %\texttt{https://media.xiph.org/video/derf/ElFuente/Netflix\_Tango\_Copyright.txt}}]}
  \label{fig:vis_gop}
\end{figure*}

\begin{figure*}[t]
  \begin{center}
    \includegraphics[width=.96\textwidth]{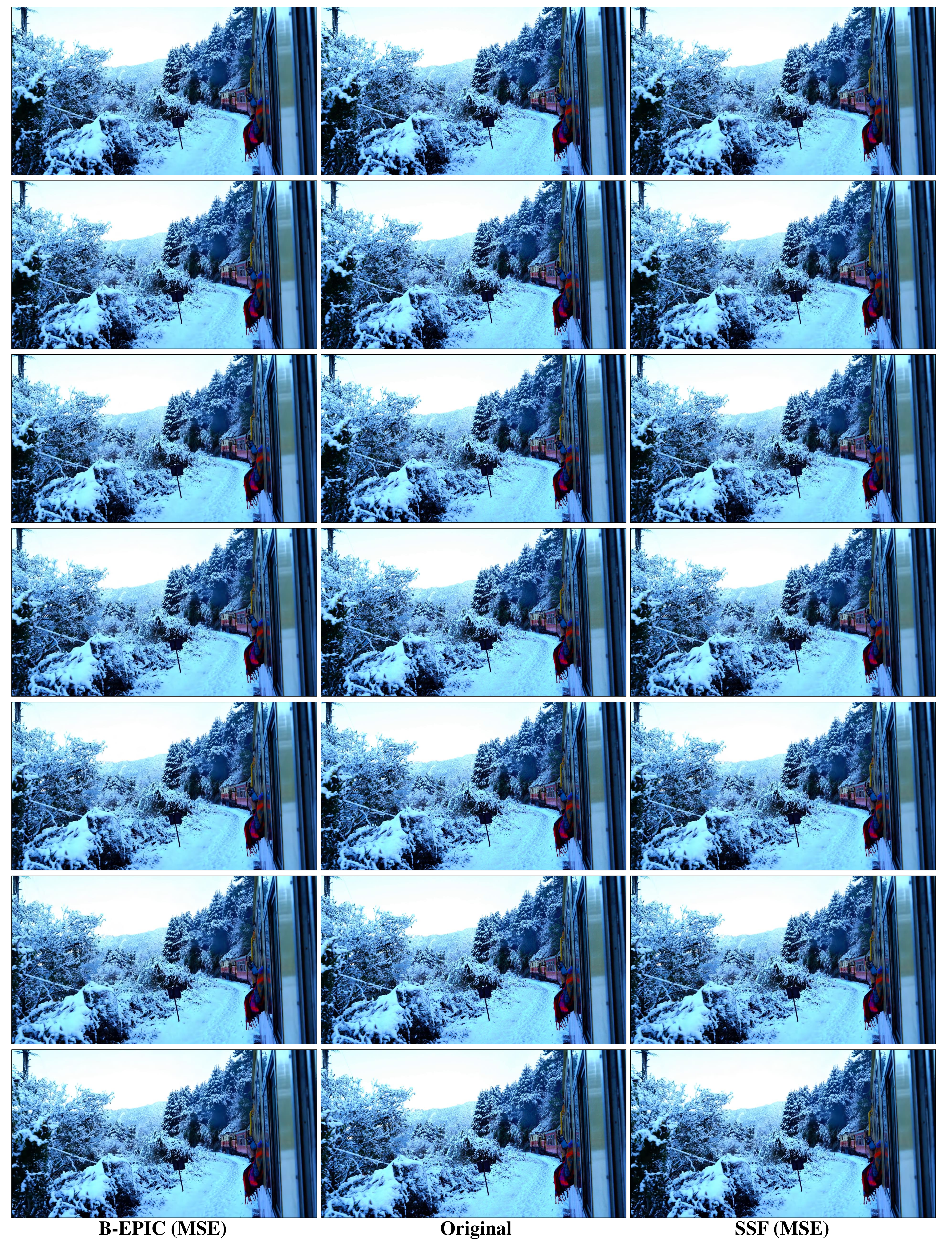}
  \end{center}
  \vspace{-1em}
  \caption{Qualitative comparisons of $\bepic$ and $\ssf$. The columns represent a sequence of 7 frames (top to bottom) compressed using $\bepic$ and $\ssf$ as well as the uncompressed sequence where the GoP structures for $\bepic$ and $\ssf$ are \texttt{IBBBBBP} and \texttt{IPPPPPP}, respectively. The average (size, $\psnr$) for $\bepic$ and $\ssf$ are (43.47KB, 28.80dB) and (51.39KB, 28.85dB). While $\bepic$ generates similar $\psnr$ and visual quality, it consumes 15.4\% less bits compared to $\ssf$. %\\
   \scriptsize{[Video obtained from \href{https://www.pexels.com/video/train-traveling-on-the-mountainside-in-winter-3563083/}{Pexels}.}]}
  %\scriptsize{[Video obtained from Pexels.}]}
  \label{fig:vis_perc_1}
\end{figure*}

\begin{figure*}[t]
  \begin{center}
    \includegraphics[width=.96\textwidth]{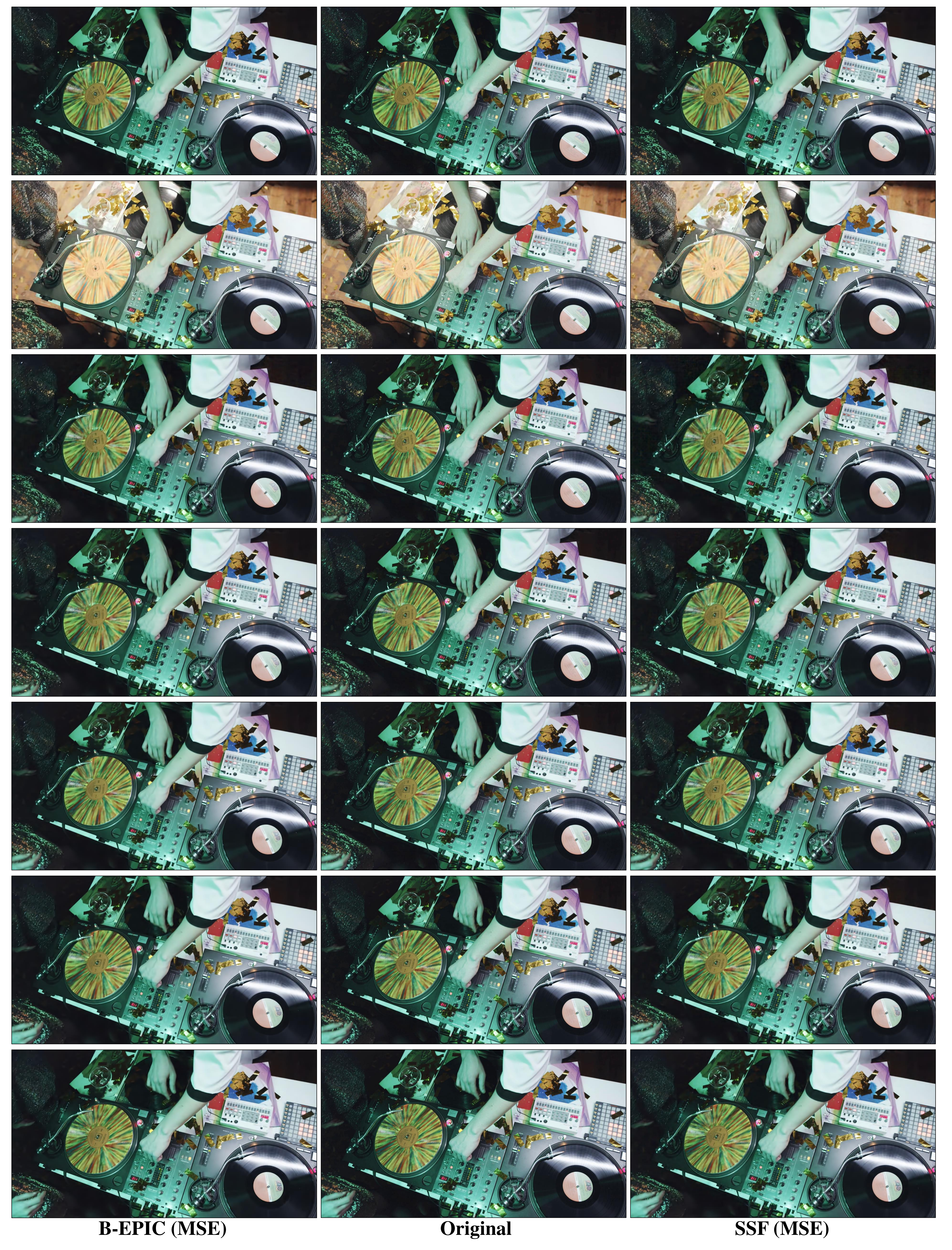}
  \end{center}
  \vspace{-1em}
  \caption{Qualitative comparisons of $\bepic$ and $\ssf$. The columns represent a sequence of 7 frames (top to bottom) compressed using $\bepic$ and $\ssf$ as well as the uncompressed sequence where the GoP structures for $\bepic$ and $\ssf$ are \texttt{IBBBBBP} and \texttt{IPPPPPP}, respectively. The average (size, $\psnr$) for $\bepic$ and $\ssf$ are (23.84KB, 33.81dB) and (27.01KB, 33.94dB). While $\bepic$ generates similar $\psnr$ and visual quality, it consumes 11.7\% less bits compared to $\ssf$. %\\
   \scriptsize{[Video obtained from \href{https://www.pexels.com/video/a-disc-jockey-at-work-3403528/}{Pexels}.}]}
  %\scriptsize{[Video obtained from Pexels.}]}
  \label{fig:vis_perc_2}
\end{figure*}

\textbf{Rate-distortion results}: 
We report the per-video performance of our $\bepicmse$ and $\bepicmsssim$ models on the UVG~\cite{UVG}, MCL-JCV~\cite{mcljcv}, and HEVC~\cite{HEVC_dataset} datasets in Tables~\ref{tab:uvg_psnr_detailed} through~\ref{tab:hevce_msssim_detailed}.

\begin{table*}[t]
    \centering
    \scriptsize
    \begin{tabular*}{\textwidth}{c|c@{\hspace{7pt}}c|c@{\hspace{7pt}}c|c@{\hspace{7pt}}c|c@{\hspace{7pt}}c|c@{\hspace{7pt}}c|c@{\hspace{7pt}}c|c@{\hspace{7pt}}c|c@{\hspace{7pt}}c}
        \hline
         & \multicolumn{16}{c}{Performance across models - $\psnr$ (dB) vs $\rate$ (bits-per-pixel)} \\
        \hline
        Video & $\rate$ & $\psnr$ & $\rate$ & $\psnr$ & $\rate$ & $\psnr$ & $\rate$ & $\psnr$ & $\rate$ & $\psnr$ & $\rate$ & $\psnr$ & $\rate$ & $\psnr$ & $\rate$ & $\psnr$ \\
        \hline
		Beauty & 0.956 & 37.92 & 0.553 & 36.8 & 0.229 & 35.53 & 0.061 & 34.51 & 0.029 & 34.16 & 0.018 & 33.9 & 0.012 & 33.54 & 0.009 & 33.1\\
		Bosphorus & 0.235 & 42.71 & 0.12 & 41.41 & 0.067 & 40.09 & 0.04 & 38.8 & 0.025 & 37.34 & 0.016 & 36.0 & 0.013 & 34.45 & 0.01 & 33.1\\
		HoneyBee & 0.415 & 39.87 & 0.112 & 38.56 & 0.036 & 37.66 & 0.017 & 36.68 & 0.01 & 35.49 & 0.007 & 34.4 & 0.007 & 33.01 & 0.006 & 31.68\\
		Jockey & 0.419 & 40.79 & 0.193 & 39.73 & 0.11 & 39.0 & 0.071 & 38.29 & 0.053 & 37.42 & 0.034 & 36.59 & 0.026 & 35.44 & 0.02 & 34.18\\
		ReadySetGo & 0.455 & 41.38 & 0.28 & 40.14 & 0.178 & 38.75 & 0.119 & 37.26 & 0.084 & 35.65 & 0.054 & 34.12 & 0.039 & 32.46 & 0.031 & 30.77\\
		ShakeNDry & 0.599 & 40.13 & 0.292 & 38.7 & 0.159 & 37.42 & 0.094 & 36.17 & 0.058 & 34.81 & 0.035 & 33.46 & 0.02 & 31.92 & 0.013 & 30.68\\
		YachtRide & 0.474 & 41.87 & 0.316 & 40.64 & 0.21 & 39.28 & 0.139 & 37.75 & 0.091 & 36.1 & 0.057 & 34.46 & 0.036 & 32.6 & 0.024 & 30.96\\
		\hline
		Average & 0.508 & 40.67 & 0.267 & 39.43 & 0.141 & 38.25 & 0.077 & 37.07 & 0.05 & 35.85 & 0.032 & 34.71 & 0.022 & 33.35 & 0.016 & 32.07\\
        \hline
    \end{tabular*}
    \caption{Detailed rate-distortion performance of $\bepicmse$ on the UVG dataset. 
    %$\rate$ and $\psnr$ are measured in terms of bits-per-pixel and dB respectively.
    }
    \label{tab:uvg_psnr_detailed}
\end{table*}

\begin{table*}[t]
    \centering
    \scriptsize
    \begin{tabular*}{\textwidth}{c|@{\hspace{2pt}}c@{\hspace{2pt}}c@{\hspace{2pt}}|@{\hspace{2pt}}c@{\hspace{2pt}}c@{\hspace{2pt}}|@{\hspace{2pt}}c@{\hspace{2pt}}c@{\hspace{2pt}}|@{\hspace{2pt}}c@{\hspace{2pt}}c@{\hspace{2pt}}|@{\hspace{2pt}}c@{\hspace{2pt}}c@{\hspace{2pt}}|@{\hspace{2pt}}c@{\hspace{2pt}}c@{\hspace{2pt}}|@{\hspace{2pt}}c@{\hspace{2pt}}c@{\hspace{2pt}}|@{\hspace{2pt}}c@{\hspace{2pt}}c}
        \hline
        & \multicolumn{16}{c}{Performance across models - $\msssim$ vs $\rate$ (bits-per-pixel)} \\
        \hline
        Video & $\rate$ & $\msssim$ & $\rate$ & $\msssim$ & $\rate$ & $\msssim$ & $\rate$ & $\msssim$ & $\rate$ & $\msssim$ & $\rate$ & $\msssim$ & $\rate$ & $\msssim$ & $\rate$ & $\msssim$ \\
        \hline
		Beauty & 0.842 & 0.98 & 0.607 & 0.973 & 0.377 & 0.961 & 0.21 & 0.946 & 0.102 & 0.928 & 0.031 & 0.906 & 0.017 & 0.898 & 0.01 & 0.892\\
		Bosphorus & 0.37 & 0.995 & 0.203 & 0.993 & 0.112 & 0.99 & 0.065 & 0.986 & 0.035 & 0.98 & 0.021 & 0.972 & 0.015 & 0.959 & 0.012 & 0.942\\
		HoneyBee & 0.47 & 0.992 & 0.253 & 0.989 & 0.105 & 0.985 & 0.054 & 0.981 & 0.01 & 0.974 & 0.016 & 0.966 & 0.011 & 0.95 & 0.008 & 0.934\\
		Jockey & 0.589 & 0.991 & 0.365 & 0.987 & 0.207 & 0.981 & 0.115 & 0.976 & 0.07 & 0.969 & 0.039 & 0.962 & 0.03 & 0.952 & 0.02 & 0.934\\
		ReadySetGo & 0.413 & 0.995 & 0.245 & 0.993 & 0.146 & 0.99 & 0.095 & 0.986 & 0.063 & 0.98 & 0.039 & 0.971 & 0.028 & 0.958 & 0.02 & 0.939\\
		ShakeNDry & 0.573 & 0.992 & 0.362 & 0.988 & 0.207 & 0.983 & 0.135 & 0.976 & 0.065 & 0.966 & 0.049 & 0.954 & 0.027 & 0.929 & 0.018 & 0.905\\
		YachtRide & 0.47 & 0.995 & 0.296 & 0.993 & 0.19 & 0.99 & 0.121 & 0.986 & 0.08 & 0.98 & 0.05 & 0.971 & 0.03 & 0.954 & 0.019 & 0.931\\
		\hline
		Average & 0.532 & 0.991 & 0.333 & 0.988 & 0.192 & 0.983 & 0.114 & 0.977 & 0.061 & 0.968 & 0.035 & 0.957 & 0.023 & 0.943 & 0.015 & 0.925\\
        \hline
    \end{tabular*}
    \caption{Detailed rate-distortion performance of $\bepicmsssim$ on the UVG dataset.
    % $\rate$ is measured in terms of bits-per-pixel.
    }
    \label{tab:uvg_msssim_detailed}
\end{table*}

\begin{table*}[h]
    \centering
    \scriptsize
    \begin{tabular*}{\textwidth}{c|c@{\hspace{7pt}}c|c@{\hspace{7pt}}c|c@{\hspace{7pt}}c|c@{\hspace{7pt}}c|c@{\hspace{7pt}}c|c@{\hspace{7pt}}c|c@{\hspace{7pt}}c|c@{\hspace{7pt}}c}
        \hline
        & \multicolumn{16}{c}{Performance across models - $\psnr$ (dB) vs $\rate$ (bits-per-pixel)} \\
        \hline
        Video & $\rate$ & $\psnr$ & $\rate$ & $\psnr$ & $\rate$ & $\psnr$ & $\rate$ & $\psnr$ & $\rate$ & $\psnr$ & $\rate$ & $\psnr$ & $\rate$ & $\psnr$ & $\rate$ & $\psnr$ \\
        \hline
		videoSRC01 & 0.371 & 40.88 & 0.093 & 39.54 & 0.036 & 38.96 & 0.017 & 38.49 & 0.011 & 37.97 & 0.007 & 37.39 & 0.007 & 36.65 & 0.006 & 35.88\\
		videoSRC02 & 0.19 & 44.26 & 0.124 & 43.36 & 0.086 & 42.36 & 0.061 & 41.23 & 0.046 & 39.94 & 0.032 & 38.64 & 0.025 & 37.09 & 0.02 & 35.36\\
		videoSRC03 & 0.328 & 41.5 & 0.166 & 40.29 & 0.09 & 39.08 & 0.05 & 37.78 & 0.029 & 36.31 & 0.018 & 35.07 & 0.013 & 33.62 & 0.01 & 32.07\\
		videoSRC04 & 0.858 & 41.28 & 0.648 & 39.92 & 0.481 & 38.46 & 0.339 & 36.71 & 0.231 & 35.07 & 0.152 & 33.55 & 0.093 & 31.88 & 0.056 & 30.36\\
		videoSRC05 & 0.905 & 37.94 & 0.467 & 36.77 & 0.266 & 35.89 & 0.169 & 35.0 & 0.113 & 34.01 & 0.074 & 33.0 & 0.051 & 31.68 & 0.038 & 30.25\\
		videoSRC06 & 1.374 & 33.86 & 1.021 & 33.42 & 0.647 & 32.51 & 0.169 & 31.15 & 0.016 & 30.63 & 0.009 & 30.54 & 0.007 & 30.43 & 0.006 & 30.3\\
		videoSRC07 & 0.895 & 37.22 & 0.55 & 36.31 & 0.253 & 35.18 & 0.111 & 34.33 & 0.061 & 33.64 & 0.038 & 32.99 & 0.025 & 32.16 & 0.017 & 31.37\\
		videoSRC08 & 0.62 & 38.96 & 0.205 & 37.69 & 0.1 & 37.1 & 0.054 & 36.5 & 0.031 & 35.8 & 0.019 & 35.03 & 0.014 & 34.14 & 0.011 & 33.2\\
		videoSRC09 & 1.107 & 38.1 & 0.744 & 36.83 & 0.446 & 35.4 & 0.233 & 33.74 & 0.128 & 32.05 & 0.078 & 30.42 & 0.05 & 28.59 & 0.035 & 26.83\\
		videoSRC10 & 0.899 & 40.18 & 0.612 & 38.8 & 0.417 & 37.32 & 0.286 & 35.81 & 0.202 & 34.29 & 0.132 & 32.77 & 0.093 & 31.08 & 0.084 & 29.35\\
		videoSRC11 & 0.274 & 44.58 & 0.186 & 43.39 & 0.126 & 42.07 & 0.086 & 40.65 & 0.061 & 39.1 & 0.039 & 37.58 & 0.028 & 36.0 & 0.022 & 34.33\\
		videoSRC12 & 0.338 & 41.55 & 0.19 & 39.79 & 0.11 & 38.04 & 0.061 & 36.18 & 0.034 & 34.47 & 0.02 & 32.94 & 0.014 & 31.55 & 0.011 & 30.18\\
		videoSRC13 & 0.781 & 39.28 & 0.544 & 38.11 & 0.347 & 36.65 & 0.176 & 34.74 & 0.067 & 32.58 & 0.024 & 30.35 & 0.012 & 27.8 & 0.009 & 25.8\\
		videoSRC14 & 0.544 & 40.75 & 0.284 & 39.46 & 0.166 & 38.37 & 0.107 & 37.26 & 0.072 & 36.05 & 0.048 & 34.8 & 0.035 & 33.38 & 0.026 & 31.89\\
		videoSRC15 & 1.0 & 38.48 & 0.63 & 37.19 & 0.329 & 35.69 & 0.139 & 33.99 & 0.054 & 32.41 & 0.03 & 31.18 & 0.02 & 29.87 & 0.014 & 28.55\\
		videoSRC16 & 0.232 & 41.51 & 0.09 & 40.68 & 0.05 & 40.14 & 0.03 & 39.49 & 0.02 & 38.75 & 0.013 & 37.94 & 0.011 & 36.98 & 0.008 & 35.81\\
		videoSRC17 & 0.525 & 40.34 & 0.244 & 39.15 & 0.152 & 38.29 & 0.097 & 37.36 & 0.065 & 36.28 & 0.043 & 35.12 & 0.029 & 33.73 & 0.021 & 32.49\\
		videoSRC18 & 0.316 & 39.3 & 0.191 & 37.82 & 0.12 & 36.89 & 0.079 & 35.19 & 0.055 & 33.49 & 0.034 & 31.83 & 0.024 & 29.98 & 0.017 & 28.45\\
		videoSRC19 & 0.517 & 41.81 & 0.333 & 40.35 & 0.219 & 38.82 & 0.145 & 37.24 & 0.096 & 35.61 & 0.061 & 34.05 & 0.041 & 32.51 & 0.03 & 31.07\\
		videoSRC20 & 0.409 & 42.06 & 0.284 & 40.84 & 0.202 & 39.6 & 0.14 & 38.08 & 0.1 & 36.57 & 0.068 & 35.13 & 0.05 & 33.66 & 0.038 & 32.13\\
		videoSRC21 & 0.172 & 45.04 & 0.111 & 44.25 & 0.075 & 43.46 & 0.052 & 42.54 & 0.036 & 41.5 & 0.025 & 40.38 & 0.019 & 39.1 & 0.015 & 37.78\\
		videoSRC22 & 0.525 & 42.84 & 0.348 & 41.5 & 0.242 & 40.23 & 0.172 & 38.91 & 0.124 & 37.49 & 0.088 & 35.95 & 0.063 & 34.29 & 0.046 & 32.58\\
		videoSRC23 & 0.317 & 43.45 & 0.186 & 42.05 & 0.113 & 40.61 & 0.072 & 39.13 & 0.047 & 37.57 & 0.031 & 36.04 & 0.022 & 34.47 & 0.016 & 32.88\\
		videoSRC24 & 0.333 & 41.48 & 0.216 & 40.51 & 0.146 & 39.68 & 0.095 & 38.53 & 0.061 & 37.07 & 0.039 & 35.76 & 0.027 & 33.95 & 0.02 & 32.22\\
		videoSRC25 & 1.084 & 36.19 & 0.809 & 35.27 & 0.594 & 34.45 & 0.411 & 33.06 & 0.27 & 31.49 & 0.174 & 30.02 & 0.107 & 28.09 & 0.067 & 26.31\\
		videoSRC26 & 0.243 & 42.55 & 0.138 & 41.65 & 0.093 & 40.85 & 0.065 & 39.94 & 0.047 & 38.89 & 0.034 & 37.8 & 0.026 & 36.48 & 0.021 & 34.9\\
		videoSRC27 & 0.447 & 42.4 & 0.299 & 40.89 & 0.2 & 39.28 & 0.132 & 37.59 & 0.086 & 35.82 & 0.054 & 34.12 & 0.034 & 32.3 & 0.023 & 30.7\\
		videoSRC28 & 0.178 & 42.72 & 0.085 & 41.74 & 0.046 & 40.74 & 0.027 & 39.63 & 0.018 & 38.44 & 0.012 & 37.16 & 0.011 & 35.8 & 0.009 & 34.39\\
		videoSRC29 & 0.071 & 45.3 & 0.035 & 44.59 & 0.021 & 43.96 & 0.013 & 43.28 & 0.009 & 42.5 & 0.007 & 41.71 & 0.006 & 40.84 & 0.005 & 39.79\\
		videoSRC30 & 0.286 & 40.02 & 0.144 & 39.26 & 0.054 & 38.36 & 0.022 & 37.52 & 0.013 & 36.72 & 0.009 & 35.94 & 0.008 & 35.06 & 0.007 & 34.08\\
		\hline
		Average & 0.538 & 40.86 & 0.333 & 39.71 & 0.208 & 38.62 & 0.12 & 37.37 & 0.073 & 36.08 & 0.047 & 34.84 & 0.032 & 33.44 & 0.024 & 32.04\\
        \hline
    \end{tabular*}
    \caption{Detailed rate-distortion performance of $\bepicmse$ on the MCL-JCV dataset.
    % $\rate$ and $\psnr$ are measured in terms of bits-per-pixel and dB respectively.
    }
    \label{tab:mcljcv_psnr_detailed}
\end{table*}

\begin{table*}[h]
    \centering
    \scriptsize
    \begin{tabular*}{\textwidth}{c|@{\hspace{2pt}}c@{\hspace{2pt}}c@{\hspace{2pt}}|@{\hspace{2pt}}c@{\hspace{2pt}}c@{\hspace{2pt}}|@{\hspace{2pt}}c@{\hspace{2pt}}c@{\hspace{2pt}}|@{\hspace{2pt}}c@{\hspace{2pt}}c@{\hspace{2pt}}|@{\hspace{2pt}}c@{\hspace{2pt}}c@{\hspace{2pt}}|@{\hspace{2pt}}c@{\hspace{2pt}}c@{\hspace{2pt}}|@{\hspace{2pt}}c@{\hspace{2pt}}c@{\hspace{2pt}}|@{\hspace{2pt}}c@{\hspace{2pt}}c}
        \hline
        & \multicolumn{16}{c}{Performance across models - $\msssim$ vs $\rate$ (bits-per-pixel)} \\
        \hline
        Video & $\rate$ & $\msssim$ & $\rate$ & $\msssim$ & $\rate$ & $\msssim$ & $\rate$ & $\msssim$ & $\rate$ & $\msssim$ & $\rate$ & $\msssim$ & $\rate$ & $\msssim$ & $\rate$ & $\msssim$ \\
        \hline
		videoSRC01 & 0.567 & 0.991 & 0.347 & 0.987 & 0.178 & 0.982 & 0.078 & 0.976 & 0.021 & 0.97 & 0.009 & 0.966 & 0.008 & 0.961 & 0.006 & 0.955\\
		videoSRC02 & 0.312 & 0.996 & 0.177 & 0.995 & 0.106 & 0.993 & 0.068 & 0.99 & 0.047 & 0.987 & 0.031 & 0.982 & 0.025 & 0.976 & 0.018 & 0.964\\
		videoSRC03 & 0.467 & 0.994 & 0.273 & 0.992 & 0.136 & 0.988 & 0.064 & 0.983 & 0.029 & 0.976 & 0.016 & 0.968 & 0.014 & 0.957 & 0.011 & 0.941\\
		videoSRC04 & 0.634 & 0.994 & 0.45 & 0.991 & 0.318 & 0.986 & 0.223 & 0.98 & 0.154 & 0.971 & 0.1 & 0.957 & 0.065 & 0.935 & 0.04 & 0.907\\
		videoSRC05 & 0.642 & 0.988 & 0.415 & 0.984 & 0.249 & 0.979 & 0.144 & 0.972 & 0.085 & 0.963 & 0.051 & 0.952 & 0.035 & 0.933 & 0.023 & 0.9\\
		videoSRC06 & 0.9 & 0.95 & 0.652 & 0.938 & 0.433 & 0.919 & 0.269 & 0.896 & 0.108 & 0.865 & 0.016 & 0.839 & 0.009 & 0.834 & 0.004 & 0.83\\
		videoSRC07 & 0.756 & 0.981 & 0.512 & 0.975 & 0.307 & 0.966 & 0.158 & 0.955 & 0.08 & 0.944 & 0.043 & 0.932 & 0.029 & 0.917 & 0.019 & 0.902\\
		videoSRC08 & 0.708 & 0.986 & 0.487 & 0.982 & 0.272 & 0.975 & 0.117 & 0.967 & 0.041 & 0.96 & 0.018 & 0.953 & 0.013 & 0.946 & 0.009 & 0.934\\
		videoSRC09 & 0.62 & 0.994 & 0.371 & 0.991 & 0.211 & 0.985 & 0.125 & 0.978 & 0.067 & 0.966 & 0.041 & 0.948 & 0.029 & 0.922 & 0.021 & 0.887\\
		videoSRC10 & 0.58 & 0.995 & 0.388 & 0.993 & 0.263 & 0.99 & 0.171 & 0.985 & 0.114 & 0.979 & 0.071 & 0.971 & 0.05 & 0.96 & 0.038 & 0.945\\
		videoSRC11 & 0.29 & 0.996 & 0.175 & 0.995 & 0.114 & 0.993 & 0.077 & 0.991 & 0.054 & 0.987 & 0.034 & 0.982 & 0.026 & 0.975 & 0.018 & 0.963\\
		videoSRC12 & 0.259 & 0.995 & 0.146 & 0.992 & 0.086 & 0.988 & 0.066 & 0.983 & 0.032 & 0.973 & 0.019 & 0.961 & 0.014 & 0.945 & 0.011 & 0.926\\
		videoSRC13 & 0.338 & 0.995 & 0.158 & 0.991 & 0.072 & 0.986 & 0.036 & 0.979 & 0.011 & 0.971 & 0.006 & 0.959 & 0.006 & 0.941 & 0.007 & 0.923\\
		videoSRC14 & 0.526 & 0.994 & 0.326 & 0.991 & 0.191 & 0.987 & 0.111 & 0.983 & 0.067 & 0.976 & 0.041 & 0.968 & 0.03 & 0.956 & 0.021 & 0.938\\
		videoSRC15 & 0.727 & 0.992 & 0.501 & 0.988 & 0.315 & 0.982 & 0.174 & 0.972 & 0.075 & 0.956 & 0.039 & 0.939 & 0.018 & 0.911 & 0.013 & 0.883\\
		videoSRC16 & 0.461 & 0.992 & 0.264 & 0.988 & 0.127 & 0.984 & 0.055 & 0.98 & 0.026 & 0.976 & 0.015 & 0.971 & 0.012 & 0.965 & 0.008 & 0.949\\
		videoSRC17 & 0.604 & 0.992 & 0.385 & 0.989 & 0.227 & 0.985 & 0.133 & 0.979 & 0.078 & 0.972 & 0.051 & 0.963 & 0.034 & 0.949 & 0.022 & 0.929\\
		videoSRC18 & 0.215 & 0.994 & 0.136 & 0.991 & 0.089 & 0.986 & 0.066 & 0.98 & 0.041 & 0.969 & 0.026 & 0.954 & 0.02 & 0.932 & 0.015 & 0.897\\
		videoSRC19 & 0.496 & 0.995 & 0.326 & 0.993 & 0.213 & 0.989 & 0.139 & 0.984 & 0.089 & 0.976 & 0.055 & 0.965 & 0.037 & 0.947 & 0.025 & 0.925\\
		videoSRC20 & 0.326 & 0.996 & 0.228 & 0.994 & 0.16 & 0.991 & 0.116 & 0.987 & 0.079 & 0.98 & 0.053 & 0.97 & 0.04 & 0.956 & 0.029 & 0.931\\
		videoSRC21 & 0.261 & 0.995 & 0.146 & 0.992 & 0.085 & 0.99 & 0.053 & 0.988 & 0.036 & 0.985 & 0.024 & 0.981 & 0.019 & 0.976 & 0.013 & 0.968\\
		videoSRC22 & 0.48 & 0.995 & 0.323 & 0.993 & 0.216 & 0.99 & 0.148 & 0.986 & 0.106 & 0.981 & 0.074 & 0.974 & 0.055 & 0.962 & 0.038 & 0.944\\
		videoSRC23 & 0.334 & 0.996 & 0.196 & 0.995 & 0.113 & 0.992 & 0.068 & 0.989 & 0.04 & 0.984 & 0.027 & 0.978 & 0.019 & 0.968 & 0.014 & 0.954\\
		videoSRC24 & 0.303 & 0.996 & 0.182 & 0.995 & 0.112 & 0.992 & 0.071 & 0.99 & 0.045 & 0.986 & 0.028 & 0.98 & 0.02 & 0.969 & 0.014 & 0.946\\
		videoSRC25 & 0.637 & 0.994 & 0.447 & 0.991 & 0.302 & 0.987 & 0.207 & 0.98 & 0.135 & 0.968 & 0.085 & 0.951 & 0.054 & 0.923 & 0.033 & 0.874\\
		videoSRC26 & 0.388 & 0.994 & 0.216 & 0.991 & 0.115 & 0.988 & 0.069 & 0.985 & 0.045 & 0.981 & 0.031 & 0.977 & 0.024 & 0.969 & 0.018 & 0.956\\
		videoSRC27 & 0.374 & 0.996 & 0.238 & 0.994 & 0.158 & 0.992 & 0.107 & 0.988 & 0.066 & 0.982 & 0.043 & 0.973 & 0.027 & 0.957 & 0.018 & 0.937\\
		videoSRC28 & 0.272 & 0.994 & 0.128 & 0.992 & 0.055 & 0.99 & 0.027 & 0.988 & 0.015 & 0.985 & 0.01 & 0.981 & 0.008 & 0.975 & 0.007 & 0.965\\
		videoSRC29 & 0.19 & 0.995 & 0.089 & 0.993 & 0.042 & 0.991 & 0.02 & 0.989 & 0.012 & 0.987 & 0.007 & 0.984 & 0.007 & 0.98 & 0.006 & 0.975\\
		videoSRC30 & 0.294 & 0.989 & 0.138 & 0.985 & 0.048 & 0.981 & 0.021 & 0.976 & 0.009 & 0.971 & 0.006 & 0.965 & 0.007 & 0.957 & 0.006 & 0.946\\
		\hline
		Average & 0.465 & 0.992 & 0.294 & 0.989 & 0.177 & 0.984 & 0.106 & 0.979 & 0.06 & 0.971 & 0.036 & 0.961 & 0.025 & 0.949 & 0.017 & 0.93\\
        \hline
    \end{tabular*}
    \caption{Detailed rate-distortion performance of $\bepicmsssim$ on the MCL-JCV dataset.
    % $\rate$ is measured in terms of bits-per-pixel.
    }
    \label{tab:mcljcv_msssim_detailed}
\end{table*}

\begin{table*}[h]
    \centering
    \scriptsize
    \begin{tabular*}{\textwidth}{c|c@{\hspace{6pt}}c|c@{\hspace{6pt}}c|c@{\hspace{6pt}}c|c@{\hspace{6pt}}c|c@{\hspace{6pt}}c|c@{\hspace{6pt}}c|c@{\hspace{6pt}}c|c@{\hspace{6pt}}c}
        \hline
        & \multicolumn{16}{c}{Performance across models - $\psnr$ (dB) vs $\rate$ (bits-per-pixel)} \\
        \hline
        Video & $\rate$ & $\psnr$ & $\rate$ & $\psnr$ & $\rate$ & $\psnr$ & $\rate$ & $\psnr$ & $\rate$ & $\psnr$ & $\rate$ & $\psnr$ & $\rate$ & $\psnr$ & $\rate$ & $\psnr$ \\
        \hline
		BQTerrace & 1.18 & 36.28 & 0.799 & 35.4 & 0.414 & 34.13 & 0.185 & 32.9 & 0.089 & 31.61 & 0.046 & 30.32 & 0.026 & 28.75 & 0.018 & 27.15\\
		BasketballDrive & 0.858 & 37.98 & 0.411 & 36.76 & 0.219 & 35.85 & 0.13 & 34.93 & 0.083 & 33.93 & 0.053 & 32.91 & 0.036 & 31.55 & 0.026 & 30.15\\
		Cactus & 0.986 & 36.74 & 0.525 & 35.69 & 0.239 & 34.64 & 0.108 & 33.51 & 0.06 & 32.39 & 0.036 & 31.27 & 0.023 & 29.89 & 0.017 & 28.58\\
		Kimono & 0.55 & 39.96 & 0.247 & 38.69 & 0.138 & 37.7 & 0.082 & 36.67 & 0.052 & 35.49 & 0.034 & 34.32 & 0.023 & 32.97 & 0.017 & 31.75\\
		ParkScene & 0.805 & 38.02 & 0.445 & 36.73 & 0.241 & 35.39 & 0.122 & 33.83 & 0.061 & 32.2 & 0.034 & 30.78 & 0.021 & 29.33 & 0.016 & 28.11\\
		\hline
		Average & 0.876 & 37.8 & 0.486 & 36.66 & 0.25 & 35.54 & 0.125 & 34.37 & 0.069 & 33.13 & 0.041 & 31.92 & 0.026 & 30.5 & 0.019 & 29.15\\
        \hline
    \end{tabular*}
    \caption{Detailed rate-distortion performance of $\bepicmse$ on the HEVC class-B dataset.
    % $\rate$ and $\psnr$ are measured in terms of bits-per-pixel and dB respectively.
    }
    \label{tab:hevcb_psnr_detailed}
\end{table*}

\begin{table*}[h]
    \centering
    \scriptsize
    \begin{tabular*}{\textwidth}{c@{\hspace{1pt}}|@{\hspace{1pt}}c@{\hspace{2pt}}c@{\hspace{2pt}}|@{\hspace{2pt}}c@{\hspace{2pt}}c@{\hspace{2pt}}|@{\hspace{2pt}}c@{\hspace{2pt}}c@{\hspace{2pt}}|@{\hspace{2pt}}c@{\hspace{2pt}}c@{\hspace{2pt}}|@{\hspace{2pt}}c@{\hspace{2pt}}c@{\hspace{2pt}}|@{\hspace{2pt}}c@{\hspace{2pt}}c@{\hspace{2pt}}|@{\hspace{2pt}}c@{\hspace{2pt}}c@{\hspace{2pt}}|@{\hspace{2pt}}c@{\hspace{2pt}}c}
        \hline
        & \multicolumn{16}{c}{Performance across models - $\msssim$ vs $\rate$ (bits-per-pixel)} \\
        \hline
        Video & $\rate$ & $\msssim$ & $\rate$ & $\msssim$ & $\rate$ & $\msssim$ & $\rate$ & $\msssim$ & $\rate$ & $\msssim$ & $\rate$ & $\msssim$ & $\rate$ & $\msssim$ & $\rate$ & $\msssim$ \\
        \hline
		BQTerrace & 0.635 & 0.987 & 0.413 & 0.983 & 0.238 & 0.977 & 0.12 & 0.968 & 0.049 & 0.956 & 0.026 & 0.94 & 0.014 & 0.911 & 0.01 & 0.881 \\
		BasketballDrive & 0.608 & 0.989 & 0.381 & 0.985 & 0.22 & 0.98 & 0.121 & 0.974 & 0.065 & 0.964 & 0.038 & 0.952 & 0.026 & 0.933 & 0.017 & 0.896 \\
		Cactus & 0.7 & 0.987 & 0.452 & 0.982 & 0.243 & 0.975 & 0.122 & 0.965 & 0.048 & 0.952 & 0.028 & 0.939 & 0.018 & 0.917 & 0.013 & 0.893 \\
		Kimono & 0.587 & 0.992 & 0.371 & 0.988 & 0.211 & 0.983 & 0.119 & 0.977 & 0.063 & 0.968 & 0.04 & 0.958 & 0.026 & 0.941 & 0.017 & 0.92 \\
		ParkScene & 0.628 & 0.991 & 0.416 & 0.986 & 0.244 & 0.98 & 0.139 & 0.971 & 0.064 & 0.956 & 0.036 & 0.936 & 0.018 & 0.904 & 0.013 & 0.872 \\
		\hline
		average & 0.632 & 0.989 & 0.407 & 0.985 & 0.231 & 0.979 & 0.124 & 0.971 & 0.058 & 0.959 & 0.033 & 0.945 & 0.02 & 0.921 & 0.014 & 0.892 \\
        \hline
    \end{tabular*}
    \caption{Detailed rate-distortion performance of $\bepicmsssim$ on the HEVC class-B dataset.
    % $\rate$ is measured in terms of bits-per-pixel.
    }
    \label{tab:hevcb_msssim_detailed}
\end{table*}

\begin{table*}[h]
    \centering
    \scriptsize
    \begin{tabular*}{\textwidth}{c|c@{\hspace{6pt}}c|c@{\hspace{6pt}}c|c@{\hspace{6pt}}c|c@{\hspace{6pt}}c|c@{\hspace{6pt}}c|c@{\hspace{6pt}}c|c@{\hspace{6pt}}c|c@{\hspace{6pt}}c}
        \hline
        & \multicolumn{16}{c}{Performance across models - $\psnr$ (dB) vs $\rate$ (bits-per-pixel)} \\
        \hline
        Video & $\rate$ & $\psnr$ & $\rate$ & $\psnr$ & $\rate$ & $\psnr$ & $\rate$ & $\psnr$ & $\rate$ & $\psnr$ & $\rate$ & $\psnr$ & $\rate$ & $\psnr$ & $\rate$ & $\psnr$ \\
        \hline
		BQMall & 0.741 & 38.11 & 0.436 & 36.97 & 0.269 & 35.83 & 0.168 & 34.33 & 0.105 & 32.65 & 0.063 & 31.06 & 0.04 & 29.24 & 0.027 & 27.43\\
		BaskehallDrill & 0.68 & 36.6 & 0.392 & 35.48 & 0.257 & 34.4 & 0.167 & 33.06 & 0.106 & 31.63 & 0.066 & 30.35 & 0.042 & 28.69 & 0.029 & 27.01\\
		PartyScene & 1.182 & 33.32 & 0.799 & 32.6 & 0.546 & 31.72 & 0.354 & 30.27 & 0.202 & 28.38 & 0.107 & 26.61 & 0.054 & 24.58 & 0.031 & 22.91\\
		RaceHorses & 1.175 & 36.33 & 0.834 & 35.44 & 0.589 & 34.43 & 0.395 & 32.96 & 0.249 & 31.27 & 0.145 & 29.7 & 0.076 & 27.86 & 0.046 & 26.37\\
		\hline
		Average & 0.945 & 36.09 & 0.615 & 35.12 & 0.415 & 34.09 & 0.271 & 32.65 & 0.165 & 30.98 & 0.095 & 29.43 & 0.053 & 27.59 & 0.033 & 25.93\\
        \hline
    \end{tabular*}
    \caption{Detailed rate-distortion performance of $\bepicmse$ on the HEVC class-C dataset.
    % $\rate$ and $\psnr$ are measured in terms of bits-per-pixel and dB respectively.
    }
    \label{tab:hevcc_psnr_detailed}
\end{table*}

\begin{table*}[h]
    \centering
    \scriptsize
    \begin{tabular*}{\textwidth}{c@{\hspace{2pt}}|@{\hspace{2pt}}c@{\hspace{2pt}}c@{\hspace{2pt}}|@{\hspace{2pt}}c@{\hspace{2pt}}c@{\hspace{2pt}}|@{\hspace{2pt}}c@{\hspace{2pt}}c@{\hspace{2pt}}|@{\hspace{2pt}}c@{\hspace{2pt}}c@{\hspace{2pt}}|@{\hspace{2pt}}c@{\hspace{2pt}}c@{\hspace{2pt}}|@{\hspace{2pt}}c@{\hspace{2pt}}c@{\hspace{2pt}}|@{\hspace{2pt}}c@{\hspace{2pt}}c@{\hspace{2pt}}|@{\hspace{2pt}}c@{\hspace{2pt}}c}
        \hline
        & \multicolumn{16}{c}{Performance across models - $\msssim$ vs $\rate$ (bits-per-pixel)} \\
        \hline
        Video & $\rate$ & $\msssim$ & $\rate$ & $\msssim$ & $\rate$ & $\msssim$ & $\rate$ & $\msssim$ & $\rate$ & $\msssim$ & $\rate$ & $\msssim$ & $\rate$ & $\msssim$ & $\rate$ & $\msssim$ \\
        \hline
		BQMall & 0.439 & 0.993 & 0.235 & 0.989 & 0.13 & 0.985 & 0.077 & 0.978 & 0.045 & 0.968 & 0.025 & 0.953 & 0.018 & 0.932 & 0.014 & 0.902 \\
		BasketballDrill & 0.451 & 0.991 & 0.275 & 0.987 & 0.162 & 0.98 & 0.096 & 0.968 & 0.055 & 0.951 & 0.033 & 0.928 & 0.025 & 0.899 & 0.017 & 0.853 \\
		PartyScene & 0.474 & 0.99 & 0.288 & 0.986 & 0.165 & 0.979 & 0.098 & 0.965 & 0.053 & 0.942 & 0.03 & 0.904 & 0.021 & 0.855 & 0.014 & 0.796 \\
		RaceHorses & 0.758 & 0.991 & 0.524 & 0.987 & 0.35 & 0.981 & 0.231 & 0.973 & 0.138 & 0.957 & 0.075 & 0.932 & 0.041 & 0.893 & 0.025 & 0.843 \\
		\hline
		average & 0.531 & 0.991 & 0.33 & 0.987 & 0.202 & 0.981 & 0.125 & 0.971 & 0.073 & 0.954 & 0.041 & 0.93 & 0.026 & 0.895 & 0.017 & 0.849 \\
        \hline
    \end{tabular*}
    \caption{Detailed rate-distortion performance of $\bepicmsssim$ on the HEVC class-C dataset.
    % $\rate$ is measured in terms of bits-per-pixel.
    }
    \label{tab:hevcc_msssim_detailed}
\end{table*}

\begin{table*}[h]
    \centering
    \scriptsize
    \begin{tabular*}{\textwidth}{c|c@{\hspace{5pt}}c|c@{\hspace{5pt}}c|c@{\hspace{5pt}}c|c@{\hspace{5pt}}c|c@{\hspace{5pt}}c|c@{\hspace{5pt}}c|c@{\hspace{5pt}}c|c@{\hspace{5pt}}c}
        \hline
        & \multicolumn{16}{c}{Performance across models - $\psnr$ (dB) vs $\rate$ (bits-per-pixel)} \\
        \hline
        Video & $\rate$ & $\psnr$ & $\rate$ & $\psnr$ & $\rate$ & $\psnr$ & $\rate$ & $\psnr$ & $\rate$ & $\psnr$ & $\rate$ & $\psnr$ & $\rate$ & $\psnr$ & $\rate$ & $\psnr$ \\
        \hline
		BQSquare & 1.249 & 34.09 & 0.87 & 33.36 & 0.593 & 32.21 & 0.391 & 30.58 & 0.217 & 28.49 & 0.109 & 26.59 & 0.043 & 24.38 & 0.02 & 22.24\\
		BasketballPass & 0.883 & 36.99 & 0.623 & 35.85 & 0.449 & 34.74 & 0.308 & 33.21 & 0.197 & 31.45 & 0.121 & 29.88 & 0.072 & 28.1 & 0.047 & 26.44\\
		BlowingBubbles & 1.124 & 33.82 & 0.729 & 32.92 & 0.468 & 31.96 & 0.281 & 30.46 & 0.154 & 28.63 & 0.08 & 26.95 & 0.042 & 25.07 & 0.025 & 23.47\\
		RaceHorses & 1.253 & 36.63 & 0.903 & 35.55 & 0.648 & 34.3 & 0.434 & 32.68 & 0.268 & 30.81 & 0.154 & 29.07 & 0.083 & 27.19 & 0.053 & 25.61\\
		\hline
		Average & 1.127 & 35.38 & 0.781 & 34.42 & 0.539 & 33.31 & 0.354 & 31.73 & 0.209 & 29.85 & 0.116 & 28.12 & 0.06 & 26.19 & 0.036 & 24.44\\
        \hline
    \end{tabular*}
    \caption{Detailed rate-distortion performance of $\bepicmse$ on the HEVC class-D dataset.
    % $\rate$ and $\psnr$ are measured in terms of bits-per-pixel and dB respectively.
    }
    \label{tab:hevcd_psnr_detailed}
\end{table*}

\begin{table*}[h]
    \centering
    \scriptsize
    \begin{tabular*}{\textwidth}{c@{\hspace{1pt}}|@{\hspace{1pt}}c@{\hspace{2pt}}c@{\hspace{2pt}}|@{\hspace{2pt}}c@{\hspace{2pt}}c@{\hspace{2pt}}|@{\hspace{2pt}}c@{\hspace{2pt}}c@{\hspace{2pt}}|@{\hspace{2pt}}c@{\hspace{2pt}}c@{\hspace{2pt}}|@{\hspace{2pt}}c@{\hspace{2pt}}c@{\hspace{2pt}}|@{\hspace{2pt}}c@{\hspace{2pt}}c@{\hspace{2pt}}|@{\hspace{2pt}}c@{\hspace{2pt}}c@{\hspace{2pt}}|@{\hspace{2pt}}c@{\hspace{2pt}}c}
        \hline
        & \multicolumn{16}{c}{Performance across models - $\msssim$ vs $\rate$ (bits-per-pixel)} \\
        \hline
        Video & $\rate$ & $\msssim$ & $\rate$ & $\msssim$ & $\rate$ & $\msssim$ & $\rate$ & $\msssim$ & $\rate$ & $\msssim$ & $\rate$ & $\msssim$ & $\rate$ & $\msssim$ & $\rate$ & $\msssim$ \\
        \hline
		BQSquare & 0.4 & 0.991 & 0.228 & 0.987 & 0.121 & 0.981 & 0.066 & 0.972 & 0.029 & 0.952 & 0.011 & 0.921 & 0.008 & 0.889 & 0.008 & 0.851 \\
		BasketballPass & 0.477 & 0.993 & 0.311 & 0.99 & 0.208 & 0.984 & 0.137 & 0.975 & 0.084 & 0.961 & 0.05 & 0.939 & 0.033 & 0.908 & 0.021 & 0.857 \\
		BlowingBubbles & 0.393 & 0.989 & 0.22 & 0.984 & 0.117 & 0.975 & 0.066 & 0.96 & 0.035 & 0.936 & 0.019 & 0.897 & 0.015 & 0.851 & 0.01 & 0.797 \\
		RaceHorses & 0.697 & 0.992 & 0.474 & 0.989 & 0.317 & 0.983 & 0.206 & 0.973 & 0.125 & 0.956 & 0.064 & 0.928 & 0.037 & 0.892 & 0.022 & 0.835 \\
		\hline
		average & 0.492 & 0.992 & 0.308 & 0.987 & 0.191 & 0.981 & 0.119 & 0.97 & 0.068 & 0.951 & 0.036 & 0.921 & 0.023 & 0.885 & 0.015 & 0.835 \\
        \hline
    \end{tabular*}
    \caption{Detailed rate-distortion performance of $\bepicmsssim$ on the HEVC class-D dataset.
    % $\rate$ is measured in terms of bits-per-pixel.
    }
    \label{tab:hevcd_msssim_detailed}
\end{table*}

\begin{table*}[h]
    \centering
    \scriptsize
    \begin{tabular*}{\textwidth}{c|c@{\hspace{8pt}}c|c@{\hspace{8pt}}c|c@{\hspace{8pt}}c|c@{\hspace{8pt}}c|c@{\hspace{8pt}}c|c@{\hspace{8pt}}c|c@{\hspace{8pt}}c|c@{\hspace{8pt}}c}
        \hline
        & \multicolumn{16}{c}{Performance across models - $\psnr$ (dB) vs $\rate$ (bits-per-pixel)} \\
        \hline
        Video & $\rate$ & $\psnr$ & $\rate$ & $\psnr$ & $\rate$ & $\psnr$ & $\rate$ & $\psnr$ & $\rate$ & $\psnr$ & $\rate$ & $\psnr$ & $\rate$ & $\psnr$ & $\rate$ & $\psnr$ \\
        \hline
		Vidyo1 & 0.234 & 42.17 & 0.108 & 40.84 & 0.052 & 39.65 & 0.029 & 38.44 & 0.018 & 37.06 & 0.012 & 35.68 & 0.01 & 33.95 & 0.008 & 32.24\\
		Vidyo3 & 0.243 & 42.48 & 0.125 & 41.17 & 0.066 & 39.81 & 0.037 & 38.46 & 0.022 & 36.78 & 0.014 & 35.08 & 0.011 & 33.1 & 0.009 & 31.24\\
		Vidyo4 & 0.241 & 42.61 & 0.118 & 41.17 & 0.058 & 39.73 & 0.032 & 38.36 & 0.019 & 36.85 & 0.012 & 35.39 & 0.01 & 33.68 & 0.008 & 32.12\\
		\hline
		Average & 0.239 & 42.42 & 0.117 & 41.06 & 0.059 & 39.73 & 0.033 & 38.42 & 0.02 & 36.89 & 0.013 & 35.39 & 0.01 & 33.58 & 0.009 & 31.87\\
        \hline
    \end{tabular*}
    \caption{Detailed rate-distortion performance of $\bepicmse$ on the HEVC class-E dataset.
    % $\rate$ and $\psnr$ are measured in terms of bits-per-pixel and dB respectively.
    }
    \label{tab:hevce_psnr_detailed}
\end{table*}

\begin{table*}[h]
    \centering
    \scriptsize
    \begin{tabular*}{\textwidth}{c@{\hspace{2pt}}|@{\hspace{2pt}}c@{\hspace{3pt}}c@{\hspace{3pt}}|@{\hspace{3pt}}c@{\hspace{3pt}}c@{\hspace{3pt}}|@{\hspace{3pt}}c@{\hspace{3pt}}c@{\hspace{3pt}}|@{\hspace{3pt}}c@{\hspace{3pt}}c@{\hspace{3pt}}|@{\hspace{3pt}}c@{\hspace{3pt}}c@{\hspace{3pt}}|@{\hspace{3pt}}c@{\hspace{3pt}}c@{\hspace{3pt}}|@{\hspace{3pt}}c@{\hspace{3pt}}c@{\hspace{3pt}}|@{\hspace{3pt}}c@{\hspace{3pt}}c}
        \hline
        & \multicolumn{16}{c}{Performance across models - $\msssim$ vs $\rate$ (bits-per-pixel)} \\
        \hline
        Video & $\rate$ & $\msssim$ & $\rate$ & $\msssim$ & $\rate$ & $\msssim$ & $\rate$ & $\msssim$ & $\rate$ & $\msssim$ & $\rate$ & $\msssim$ & $\rate$ & $\msssim$ & $\rate$ & $\msssim$ \\
        \hline
		Vidyo1 & 0.298 & 0.994 & 0.154 & 0.992 & 0.07 & 0.988 & 0.032 & 0.984 & 0.013 & 0.979 & 0.008 & 0.972 & 0.009 & 0.962 & 0.008 & 0.95 \\
		Vidyo3 & 0.254 & 0.995 & 0.112 & 0.992 & 0.047 & 0.989 & 0.026 & 0.985 & 0.012 & 0.979 & 0.008 & 0.97 & 0.009 & 0.956 & 0.008 & 0.941 \\
		Vidyo4 & 0.243 & 0.995 & 0.118 & 0.992 & 0.054 & 0.989 & 0.031 & 0.985 & 0.014 & 0.979 & 0.009 & 0.972 & 0.009 & 0.959 & 0.008 & 0.947 \\
		\hline
		average & 0.265 & 0.995 & 0.128 & 0.992 & 0.057 & 0.989 & 0.029 & 0.985 & 0.013 & 0.979 & 0.009 & 0.971 & 0.009 & 0.959 & 0.008 & 0.946 \\
        \hline
    \end{tabular*}
    \caption{Detailed rate-distortion performance of $\bepicmsssim$ on the HEVC class-E dataset.
    % $\rate$ is measured in terms of bits-per-pixel.
    }
    \label{tab:hevce_msssim_detailed}
\end{table*}

\end{document}